%% file: MakeArxivMerged.tex
\documentclass[pre,twocolumn,superscriptaddress,preprintnumbers,amsmath,amssymb]{revtex4-1}
\usepackage{epsfig}
\usepackage{color}
\usepackage{amsmath}
\usepackage{amsfonts}
\usepackage{graphicx}
\usepackage{fancybox}
\usepackage{subfigure}
\usepackage{notes2bib}
\usepackage{fontenc}
\usepackage{amssymb}
\usepackage[version=3]{mhchem}

\usepackage{subfiles}

  \makeatletter
    \renewcommand\@make@capt@title[2]{%
     \@ifx@empty\float@link{\@firstofone}{\expandafter\href\expandafter{\float@link}}%
      {\textbf{#1}}\@caption@fignum@sep#2\quad}%
    \makeatother
   
\makeatletter 
\renewcommand{\fnum@figure}{\textbf{Figure~\thefigure}}
\makeatother

\begin{document}

\subfile{Dewetting_Pathways_Feb16_submit.tex}

%\bibliography{MakeArxivMerged}

\clearpage
\newpage

\subfile{SI-Feb16-2015-submit.tex}

\end{document}

%% file: Dewetting_Pathways_Feb16_submit.tex
%%%%%%%%%%%%%%%%%%%%%%%%%%%%%%
%\bibliographystyle{pnas}

%% For titles, only capitalize the first letter
\title{Pathways to dewetting in hydrophobic confinement}

%Author List
\author{Richard C. Remsing}
\altaffiliation[]{RCR and EX are joint first authors who contributed equally to this work}
\affiliation{Department of Chemical \& Biomolecular Engineering, University of Pennsylvania, Philadelphia, PA 19104, USA}

\author{Erte Xi}
\altaffiliation[]{RCR and EX are joint first authors who contributed equally to this work}
\affiliation{Department of Chemical \& Biomolecular Engineering, University of Pennsylvania, Philadelphia, PA 19104, USA}

\author{Srivathsan Vembanur}
\affiliation{Howard P. Isermann Department of Chemical \& Biological Engineering, and Center for Biotechnology and Interdisciplinary Studies, Rensselaer Polytechnic Institute, Troy, NY 12180, USA}

\author{Sumit Sharma}

\author{Pablo G. Debenedetti}
\affiliation{Department of Chemical \& Biological Engineering, Princeton University, Princeton, NJ 08544, USA}

\author{Shekhar Garde}
\affiliation{Howard P. Isermann Department of Chemical \& Biological Engineering, and Center for Biotechnology and Interdisciplinary Studies, Rensselaer Polytechnic Institute, Troy, NY 12180, USA}

\author{Amish J. Patel}
\email{amish.patel@seas.upenn.edu}
\affiliation{Department of Chemical \& Biomolecular Engineering, University of Pennsylvania, Philadelphia, PA 19104, USA}

\date{\today}

%\begin{article}

%%%%%%%%%%%%%%%%%%%%%%%%%%%%%%%%%%%%
\begin{abstract}
%abstract of no more than 250 words
%
Liquid water can become metastable with respect to its vapor in hydrophobic confinement.
The resulting dewetting transitions are often impeded by large kinetic barriers. 
According to macroscopic theory, such barriers arise from the free energy required to nucleate a critical vapor tube that spans the region between two hydrophobic surfaces -- tubes with smaller radii collapse, whereas larger ones grow to dry the entire confined region. 
Using extensive molecular simulations of water between two nanoscopic hydrophobic surfaces, in conjunction with advanced sampling techniques, here we show that for inter-surface separations that thermodynamically favor dewetting, the barrier to dewetting does not correspond to the formation of a (classical) critical vapor tube.
Instead, it corresponds to an abrupt transition from an isolated cavity adjacent to one of the confining surfaces to a gap-spanning vapor tube that is already larger than the critical vapor tube anticipated by macroscopic theory. 
Correspondingly, the barrier to dewetting is also smaller than the classical expectation. 
We show that the peculiar nature of water density fluctuations adjacent to extended hydrophobic surfaces -- namely, the enhanced likelihood of observing low-density fluctuations relative to Gaussian statistics -- facilitates this non-classical behavior. 
By stabilizing isolated cavities relative to vapor tubes, enhanced water density fluctuations thus stabilize novel pathways, which circumvent the classical barriers and offer diminished resistance to dewetting.
Our results thus suggest a key role for fluctuations in speeding up the kinetics of numerous phenomena ranging from Cassie-Wenzel transitions on superhydrophobic surfaces, to hydrophobically-driven biomolecular folding and assembly.
\end{abstract}
%%%%%%%%%%%%%%%%%%%%%%%%%%%%%%%%%%%%

%\keywords{capillary evaporation | fluctuations | kinetic barriers | assembly }

%\abbreviations{MD, molecular dynamics; INDUS, indirect umbrella sampling}

\maketitle
\raggedbottom

%% The first letter of the article should be drop cap: \dropcap{}
The favorable interactions between two extended hydrophobic surfaces drive numerous biomolecular and colloidal assemblies~\cite{tanford_book,kauzmann,FHS:1973,Israelachvili_book,Chandler:Nature:2005}, and have been the subject of several theoretical, computational, and experimental inquiries~\cite{Israelachvili:Pashley:1982,Christenson:Science:1988,Berard:JCP:1993,Parker:JPC:1994,wallqvist2,Lum:PRE:1997,Lum:IntJT:1998,LCW,Bolhuis:JCP:2000,Leung:PRL:2003,HuangX:PNAS:2003,urbic2006confined,Choudhury:JACS:2007,Xu:JPCB:2010,Sharma:PNAS:2012,Sharma:JPCB:2012,Ducker:PRL:2012}.
Examples include the association of small proteins to form multimeric protein complexes, of amphiphlic block copolymers, dendrimers, or proteins to form vesicular suprastructures, and of patchy colloidal particles into complex crystalline lattices~\cite{Hagan:2012:Capsid,Discher:Science:1999,Percec:Science:2010,Janus:Granick,Vargo:PNAS:2012}.
When two such hydrophobic surfaces approach each other, water between them becomes metastable with respect to its vapor at a critical separation, $d_c$, that can be quite large~\cite{Berard:JCP:1993,Parker:JPC:1994,Berne:ARPC:2009,PGD:JPCL:2011,Giovambattista:ARPC:2012}.
For nanometer-sized surfaces at ambient conditions, $d_c$ is proportional to the characteristic size of the hydrophobic object, whereas for micron-sized and larger surfaces, $d_c\sim1\ \mu$m~\cite{PGD:JPCL:2011,Giovambattista:ARPC:2012}. 
However, due to the presence of large kinetic barriers separating the metastable wet and the stable dry states, the system persists in the wet state, and a dewetting transition is triggered only at much smaller separations ($\sim1$~nm)~\cite{LCW,Berne:ARPC:2009,Ducker:PRL:2012,Giovambattista:ARPC:2012}.
%%%

\begin{figure}[ht]
\begin{center}
%\vspace{-0.15in}
\includegraphics[width=0.49\textwidth]{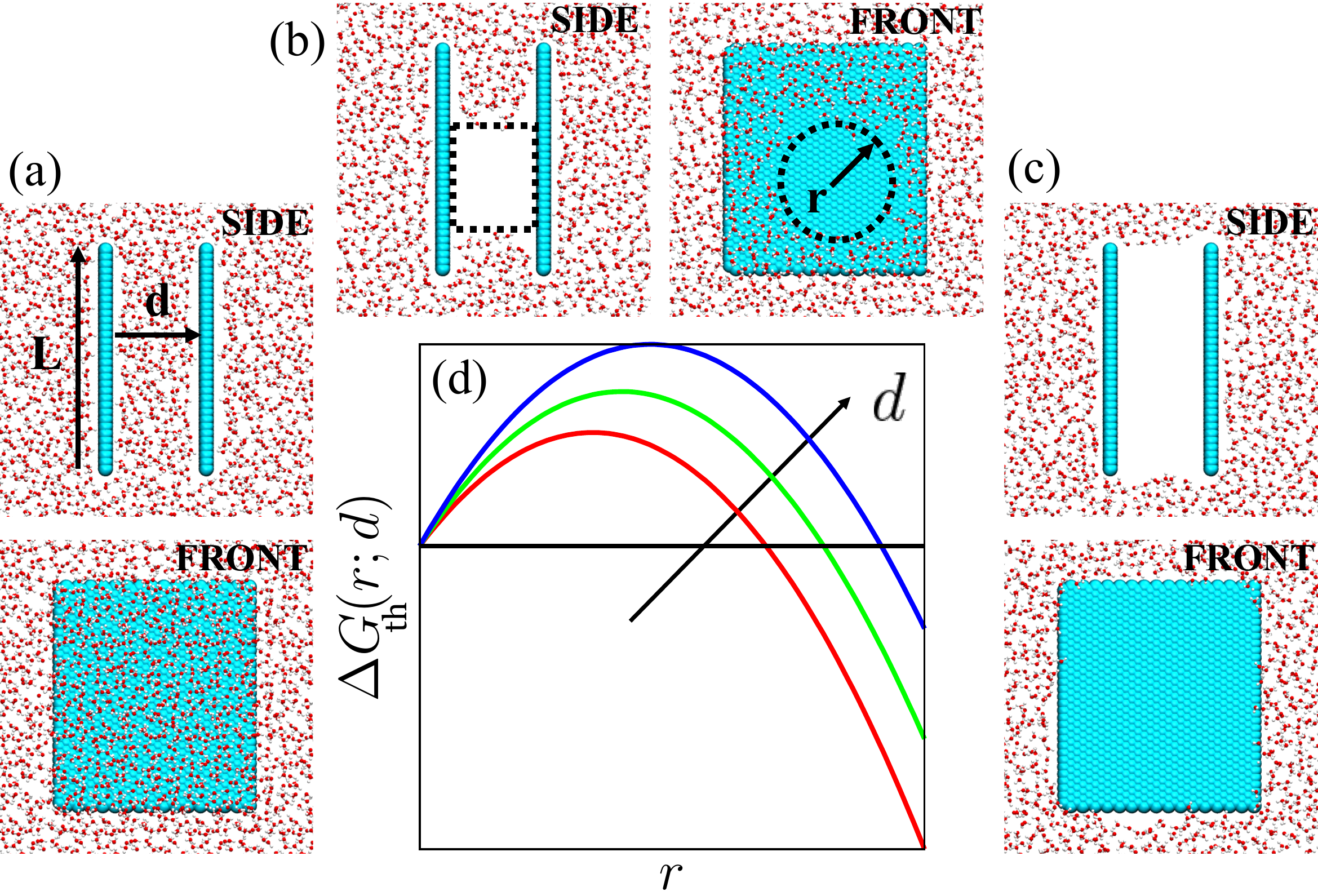}
\end{center}
\caption{
\label{fig:model}
(a - c) Simulation snapshots of water (shown in red/white) in confinement between two square hydrophobic surfaces (shown in cyan) of size $L=4$~nm that are separated by a distance of $d=20$~\AA; configurations highlighting (a) the liquid basin, (b) a cylindrical vapor tube of radius, $r$, that spans the confined region, and (c) the vapor basin are shown.
In the front views, only one of the confining surfaces is shown.
(d) Macroscopic theory predicts a free energetic barrier to vapor tube formation (Equation~\ref{eq:macro2}), suggesting that a vapor tube larger than a critical size must be nucleated before dewetting can proceed.
\vspace{-0.1in}
}
\end{figure}

%%%
To uncover the mechanism of dewetting, a number of theoretical and simulation studies have focused on the thermodynamics as well as the kinetics of dewetting in the volume between two parallel hydrophobic surfaces that are separated by a fixed distance, $d < d_c$~\cite{Berard:JCP:1993,wallqvist2,Lum:PRE:1997,Lum:IntJT:1998,LCW,Bolhuis:JCP:2000,Leung:PRL:2003,HuangX:PNAS:2003,Choudhury:JACS:2007,Xu:JPCB:2010,Sharma:PNAS:2012,Sharma:JPCB:2012}.
These studies have highlighted that the bottleneck to dewetting is the formation of a roughly cylindrical, critical vapor tube spanning the region between the surfaces~\cite{Lum:PRE:1997,Bolhuis:JCP:2000,Leung:PRL:2003}.
A barrier in the free energetics of vapor tube formation as a function of tube radius is also supported by macroscopic interfacial thermodynamics, wherein the barrier arises primarily from a competition between the favorable solid-vapor and unfavorable liquid-vapor surface energies (Equation~\ref{eq:macro2} and Figure~\ref{fig:model}).
Thus, the classical mechanism for the dewetting transition prescribes that a vapor tube, which spans the volume between the two surfaces must first be nucleated, and if the vapor tube is larger than a certain critical size, it will grow until the entire confined volume is dry~\cite{Parker:JPC:1994}.
%%%

%%%
While it has been recognized that water density fluctuations~\cite{Bolhuis:JCP:2000,Leung:PRL:2003}
 must play a crucial role in nucleating vapor tubes, the precise mechanism by which these tubes are formed is not clear.
To understand how vapor tubes are formed and to investigate their role in the dewetting process, here we use molecular simulations in conjunction with enhanced sampling methods~\cite{Patel:JPCB:2010,Patel:JSP:2011} to characterize the free energetics of water density fluctuations in the region between two nanoscopic hydrophobic surfaces.
Such a characterization of water density fluctuations in bulk water and at interfaces has already provided much insight into the physics of hydrophobic hydration and interactions~\cite{Hummer:PNAS:1996,Garde:PRL:1996,LCW,Chandler:Nature:2005,Rajamani:PNAS:2005,Mittal:PNAS:2008,Godawat:PNAS:2009,Patel:JPCB:2010,LLCW,Jamadagni:ARCB:2011,Patel:PNAS:2011,Rotenberg:JACS:2011,Patel:JSP:2011,Patel:JPCB:2012,remsing2013dissecting}.
In particular, both simulations and theory have shown that the likelihood of observing low density fluctuations adjacent to extended hydrophobic surfaces is enhanced relative to Gaussian statistics~\cite{LCW,Mittal:PNAS:2008,Godawat:PNAS:2009,Patel:JPCB:2010,LLCW,Patel:JPCB:2012}.
We show that such enhanced water density fluctuations influence the pathways to dewetting in hydrophobic confinement by stabilizing isolated cavities adjacent to one of the confining surfaces with respect to vapor tubes.
As the density in the confined region is decreased, the stability of isolated cavities relative to vapor tubes also decreases, and at a particular density, isolated cavities abruptly transition to vapor tubes.
Surprisingly, for $d \lesssim d_c$, that is, separations for which dewetting is thermodynamically favorable, we find that the nascent vapor tubes formed from the isolated cavities are already larger than the corresponding critical vapor tubes predicted by classical theory.
Because the newly formed vapor tube is super-critical, it grows spontaneously.
Importantly, because the formation of this super-critical vapor tube involves a non-classical pathway that circumvents the critical vapor tube altogether, the process entails a smaller free energetic cost.
Our results thus point to smaller kinetic barriers to dewetting than predicted by macroscopic theory.
%
%%%%%%%%%%%%%%%%%%%%%%%%%%%%%%%%%%%%

%%%%%%%%%%%%
%%%%%%%%%%%%

%%%%%%%%%%%%%%%%%%%%%%%%%%%%%%%%%%%%
\section{Macroscopic Theory} 
%%%%%%%%%%%%%%%%%%%%%%%%%%%%%%%%%%%%
%
According to classical interfacial thermodynamics, the free energy for creating a cylindrical vapor tube of radius $r$, which spans the volume between two surfaces separated by a distance $d$, is given by:
\begin{equation}
\Delta G_{\rm th}(r;d) = \pi [ r^2 d \Delta P + 2 r d \gamma + 2 r^2 \gamma \cos\theta +  4 r \lambda],
\label{eq:macro2}
\end{equation}
where $\Delta P$ is the difference between the system pressure and the saturation pressure, $\gamma$ is the liquid-vapor surface tension, $\theta$ is the contact angle, and $\lambda$ is the line tension.
For nanoscopic surfaces, the pressure-volume contribution is negligible at ambient conditions~\cite{PGD:JPCL:2011,Ashbaugh:JCP:2013,Altabet:JCP:2014}, whereas the line tension contribution can be important~\cite{Sharma:PNAS:2012,Charlaix:PNAS:2012}. 
The term containing $\cos\theta$ is negative for hydrophobic surfaces and favors dewetting, whereas the term corresponding to formation of the vapor-liquid area is unfavorable. 
The functional form of $\Delta G_{\rm th}(r;d)$ given in Equation~\ref{eq:macro2} is illustrated in Figure~\ref{fig:model}d for three $d$-values.
In each case, a barrier separates the liquid ($r=0$) and vapor (large $r$) basins, supporting the notion of dewetting mediated by the nucleation and growth of a vapor tube; both the critical vapor tube radius and the barrier height increase with increasing $d$.

%%%%%%%%%%%%
\begin{figure}[ht]
\begin{center}
\vspace{-0.15in}
\includegraphics[width=0.375\textwidth]{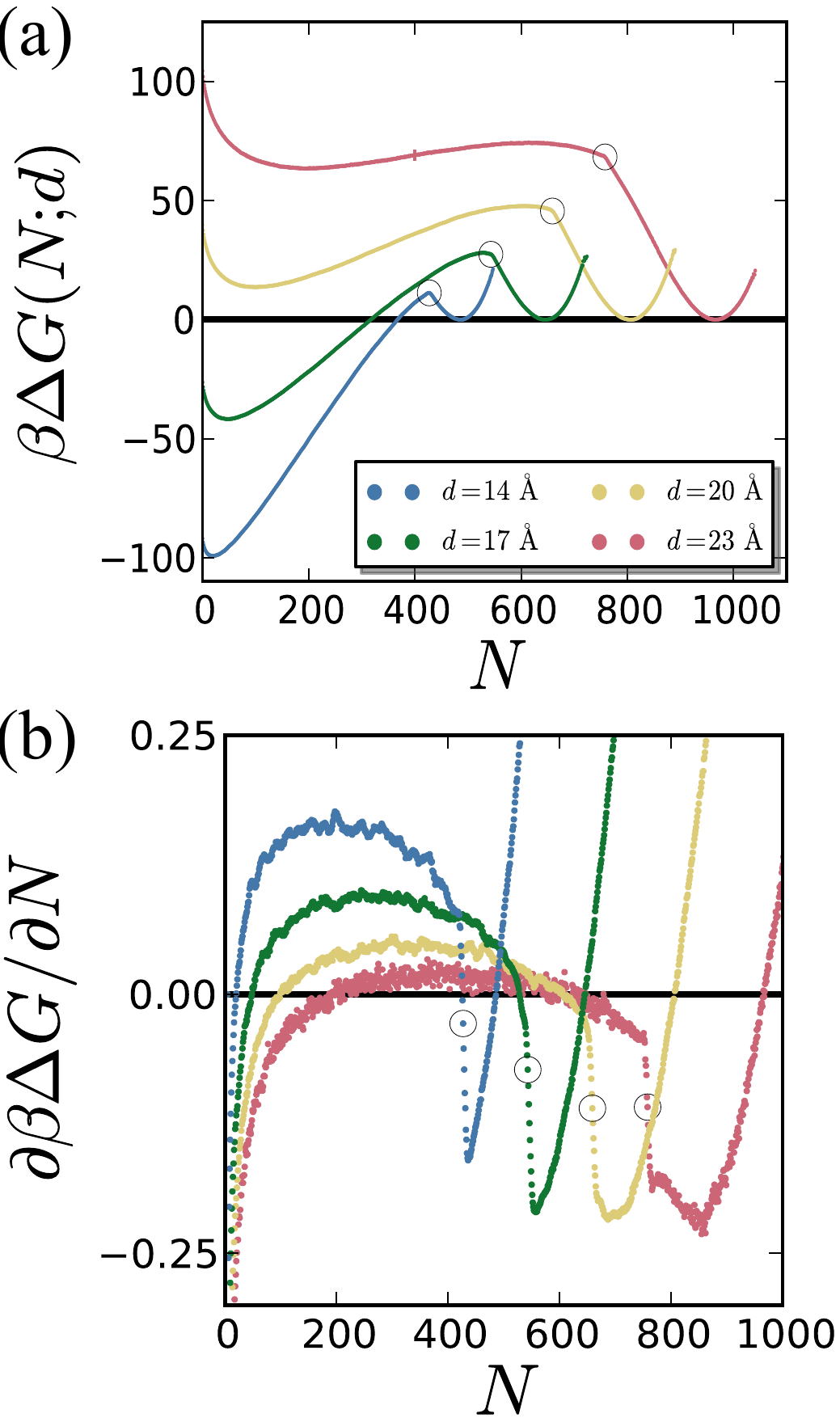}
\end{center}
\caption{
\label{fig:free}
(a) The simulated free energy profiles, $\beta\DG(N;d)$, as a function of the number of water molecules between the surfaces, $N$, display marked kinks (highlighted by circles). 
Here, $\beta=1/\kB T$, with $\kB$ being the Boltzmann constant and $T$ being the temperature.
The size of the largest error bar is also shown for $d=23$~\AA~and $N=400$.
(b) The kinks are also apparent in the smoothed derivatives of the free energy profiles, which display a sharp decrease in the vicinity of $\Nkink$. 
\vspace{-0.1in}
}
\end{figure}
%%%%%%%%%%%%

%%%%%%%%%%%%%%%%%%%%%%%%%%%%%%%%%%%%
\section{Density Dependent Free Energy Profiles Feature a Kink} 
%%%%%%%%%%%%%%%%%%%%%%%%%%%%%%%%%%%%
%
To investigate how water density fluctuations influence the mechanism of dewetting in hydrophobic confinement, here we perform molecular dynamics simulations of water confined between two, roughly square, hydrophobic surfaces of size $L=4$~nm, separated by a distance, $d$, as shown in Figure~\ref{fig:model}a.
Water in the confined region is in equilibrium with a reservoir of water, which in turn is in coexistence with its vapor ($\Delta P=0$)~\cite{Miller:PNAS:2007,Patel:JPCB:2010}.
We characterize the statistics of water density fluctuations in the confined volume using Indirect Umbrella Sampling (INDUS)~\cite{Patel:JPCB:2010,Patel:JSP:2011}, that is, we estimate the free energy, $\DG$, of observing $N$ water molecules in that volume.
The free energy, $\DG(N; d)$, thus estimated is shown in Figure~\ref{fig:free}a for a range of separations, $d$, with the free energy of the liquid basin, $N=N_{\rm liq}$, being set to zero in each case.
Over the entire range of separations considered, the free energy profile displays distinctive liquid (high $N$) and vapor (low $N$) basins with barriers separating them.
Interestingly, the free energy profiles also feature a kink, that is, an abrupt change in the slope of $\DG(N;d)$ is observed at a particular value of $N$ between the liquid and vapor basins; we refer to this value of $N$ as $\Nkink$.
This discontinuity in slope is seen more clearly in the derivatives of the free energy, shown in Figure~\ref{fig:free}b.
Small errors in $\DG$ are amplified if simple finite differences are used to evaluate the derivatives; we therefore smooth the free energy profiles before evaluating the derivatives. 
Details of the smoothing procedure as well as the unsmoothed derivatives are shown in the SI.
%

%%%%%%%%%%%%%%%%%%%%%%%%%%%%%%%%%%%%
\section{Kink Separates the Vapor Tube and Isolated Cavity Ensembles} 
\begin{figure}[t]
%\vspace{-0.1in}
\begin{center}
\includegraphics[width=0.45\textwidth]{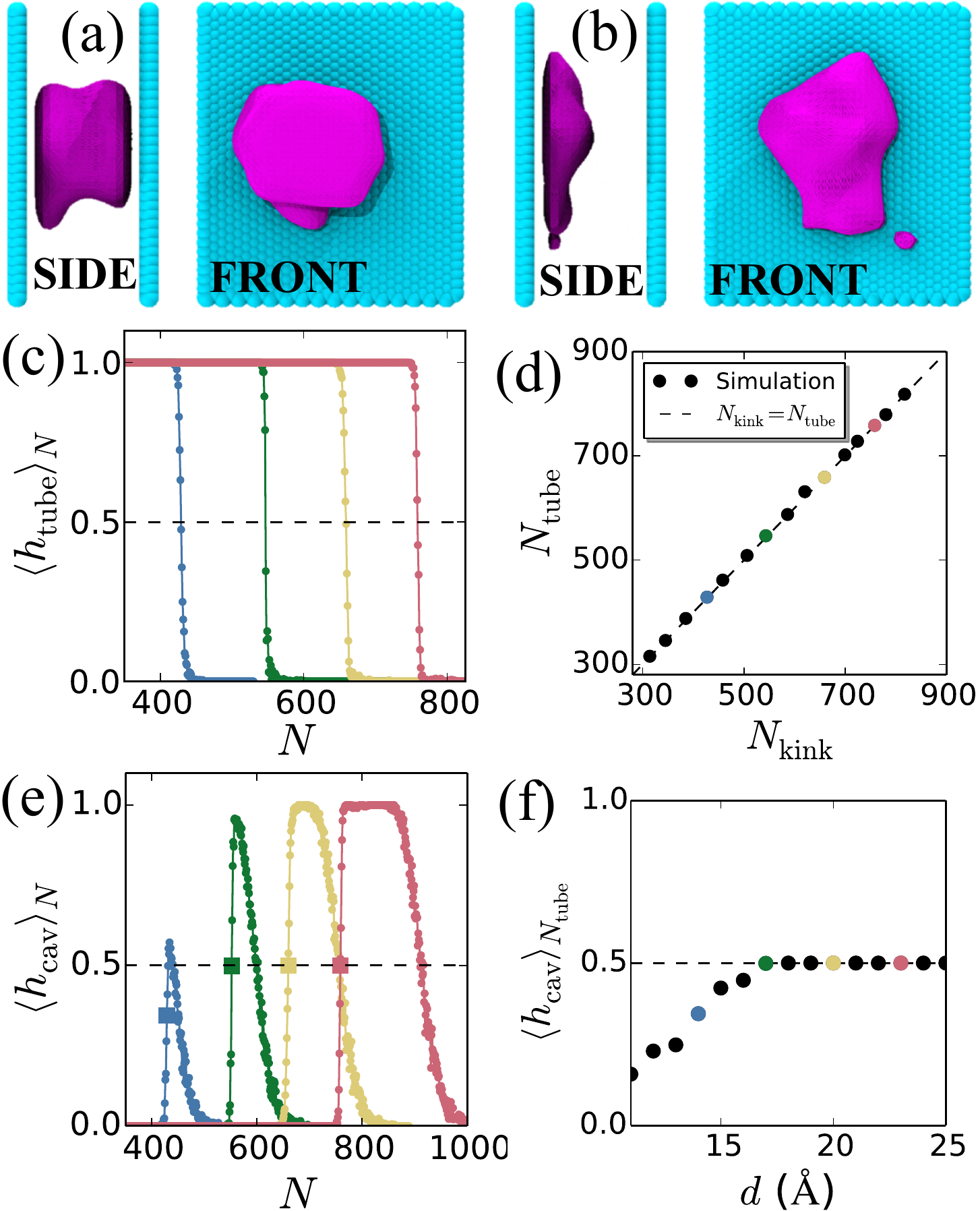}
\end{center}
%\vspace{-0.1in}
\caption{
\label{fig:kink}
Instantaneous interfaces encompassing dewetted regions (shown in purple) between the hydrophobic surfaces (shown in cyan) separated by $d=20$~\AA~highlight the presence of (a) a vapor tube for $N = \Nkink - 12$, and (b) an isolated cavity for $N = \Nkink + 3$.
Water molecules not shown for clarity. 
(c) Average of the binary vapor tube indicator function, $\htube$, conditioned on the number of waters in confinement being $N$, displays a sharp transition from 1 to 0 as $N$ is increased. 
The color scheme is the same as that in Figure~\ref{fig:free}.
The value of $N$ corresponding to $\langle\htube\rangle_N=0.5$ (dashed line) is defined as $\Ntube$.
(d) $\Ntube$ is identical to the location of the kink in the free energy profiles, $\Nkink$, confirming that the kink demarcates conformations with and without vapor tubes. 
(e) Conditional average of the isolated cavity indicator function, $\langle\hcav\rangle_N$, shows a sharp increase in the vicinity of $\Nkink$ (the square symbols correspond to $\Ntube$), followed by a gradual decrease at larger $N$-values, and eventually vanishing around $N=\Nliq$.
(f) 
For the larger $d$-values, $\langle \htube \rangle_{\Ntube} = \langle \hcav \rangle_{\Ntube} = 0.5$.
However, for the smaller $d$-values, $\langle \hcav \rangle_{\Ntube}<0.5$, suggesting the possibility of direct vapor tube nucleation without isolated cavities as intermediates.
}
\end{figure}

%%%%%%%%%%%%%%%%%%%%%%%%%%%%%%%%%%%%
%
To investigate the significance of the kink in the free energy, we characterize configurations corresponding to $N$ on either side of $\Nkink$.
We do so by building upon the instantaneous interface method of Willard and Chandler~\cite{Willard:JPCB:2010} to identify iso-surfaces that encompass the dewetted regions, that is, the regions from which water is absent.
The details of the method are included in the SI.
As illustrated in Figure~\ref{fig:kink}a for $d=20$~\AA~and $N=\Nkink-12$, characteristic configurations with $N\lesssim\Nkink$ contain a vapor tube, that is, the dewetted region (in purple) clearly spans the confined volume between the two surfaces (side view, left).
Water molecules (not shown for clarity) occupy the entire region between the surfaces not shown in purple and are also present outside the confinement region.
In contrast, for configurations with $N\gtrsim\Nkink$, isolated cavities are observed adjacent to one surface or the other; however, as seen in Figure~\ref{fig:kink}b for $N=\Nkink+3$, the cavities do not span the region between the surfaces to form vapor tubes.
Movies corresponding to the configurations shown in Figures~\ref{fig:kink}a and~\ref{fig:kink}b can be found in the SI.
%%%

%%%%%%%%%%%%
%%%%%%%%%%%%

%%%
The configurations shown in Figures~\ref{fig:kink}a and~\ref{fig:kink}b suggest that $N=\Nkink$ marks the boundary between the vapor tube and isolated cavity ensembles. 
To put this notion on a quantitative footing, we define indicator functions, $\htube$ and $\hcav$, which are 1 if a given configuration has a vapor tube or an isolated cavity respectively, and 0 otherwise.
The average value of the indicator function $\langle\htube\rangle_N$, subject to the constraint that the number of water molecules in confinement is $N$, is shown in Figure~\ref{fig:kink}c for the entire range of $N$-values, and for several separations, $d$.
Because the configurations were generated in the presence of biasing potentials, care must be exercised in evaluating the averages shown in Figure~\ref{fig:kink}c; details of the averaging procedure as well as the criteria employed in the definition of the indicator functions can be found in the SI.
For a given separation, $\langle\htube\rangle_N$, which is the probability of observing a vapor tube conditional on the number of waters in confinement being $N$, is a sigmoidal function of $N$, decreasing sharply from 1 at low $N$ to 0 for high $N$.
We define $\Ntube$ to be the value of $N$ at which $\langle\htube\rangle_N$ undergoes a sharp transition; in particular, where $\langle\htube\rangle_N$ crosses 0.5.
As shown in Figure~\ref{fig:kink}d, for the entire range of $d$-values studied here ($11\ \angstrom\le d \le 25\ \angstrom$), $\Ntube$ is equal to $\Nkink$, formalizing the notion that the kink in $\DG(N;d)$ demarcates configurations that display vapor tubes and those that do not.
%
%%%

%%%
%
Analogous to $\langle\htube\rangle_N$, the conditional average $\langle\hcav\rangle_N$ quantifies the fraction of configurations with $N$ confined waters, which feature isolated cavities.
For all separations, $\langle\hcav\rangle_N$ displays a sharp increase, followed by a gradual decrease; see Figure~\ref{fig:kink}e.
The sharp increase occurs in the vicinity of $N=\Nkink$ as vapor tubes give way to isolated cavities, whereas the gradual decrease corresponds to a crossover from isolated cavities to uniform configurations.
To specify the location of this crossover, we define $\Ncav$ to be the value of $N$ where $\langle\hcav\rangle_N$ crosses 0.5 (with a negative slope).
Despite these common features in the functional form of $\langle\hcav\rangle_N$, there are subtle differences in $\langle\hcav\rangle_N$ at small and large separations, which nevertheless have interesting consequences.
For the largest separations, a well-defined plateau at $\langle\hcav\rangle_N=1$ separates the sharp increase in $\langle\hcav\rangle_N$ and its gradual decrease; this plateau demarcates the range of $N$-values, which reliably feature isolated cavities.
In contrast, the plateau at $\langle\hcav\rangle_N=1$ is absent for the smaller separations.
Instead, the sharp increase in $\langle\hcav\rangle_N$ and its gradual decrease overlap in their range of $N$-values.
This overlap suggests that configurations corresponding to $N=\Nkink$ are not limited to those with vapor tubes or isolated cavities, but could also be uniform.
Given that $\langle\htube\rangle_N=0.5$ at $N=\Ntube$ by definition, a value of $\langle\hcav\rangle_\Ntube<0.5$ would correspond to a non-zero likelihood of observing uniform configurations with $N=\Ntube=\Nkink$.
As shown in Figure~\ref{fig:kink}f, that is indeed the case $d\le16$~\AA, suggesting that while the nucleation of a vapor tube must proceed through the formation of isolated cavities for $d>16$~\AA, vapor tubes may be nucleated directly from uniform configurations for smaller separations.
%%%

%%%%%%%%%%%%%%%%%%%%%%%%%%%%%%%%%%%%
\section{Free Energetics of Vapor Tubes and Isolated Cavities} 
\begin{figure}[t]
%\vspace{-0.25in}
\begin{center}
\includegraphics[width=0.4\textwidth]{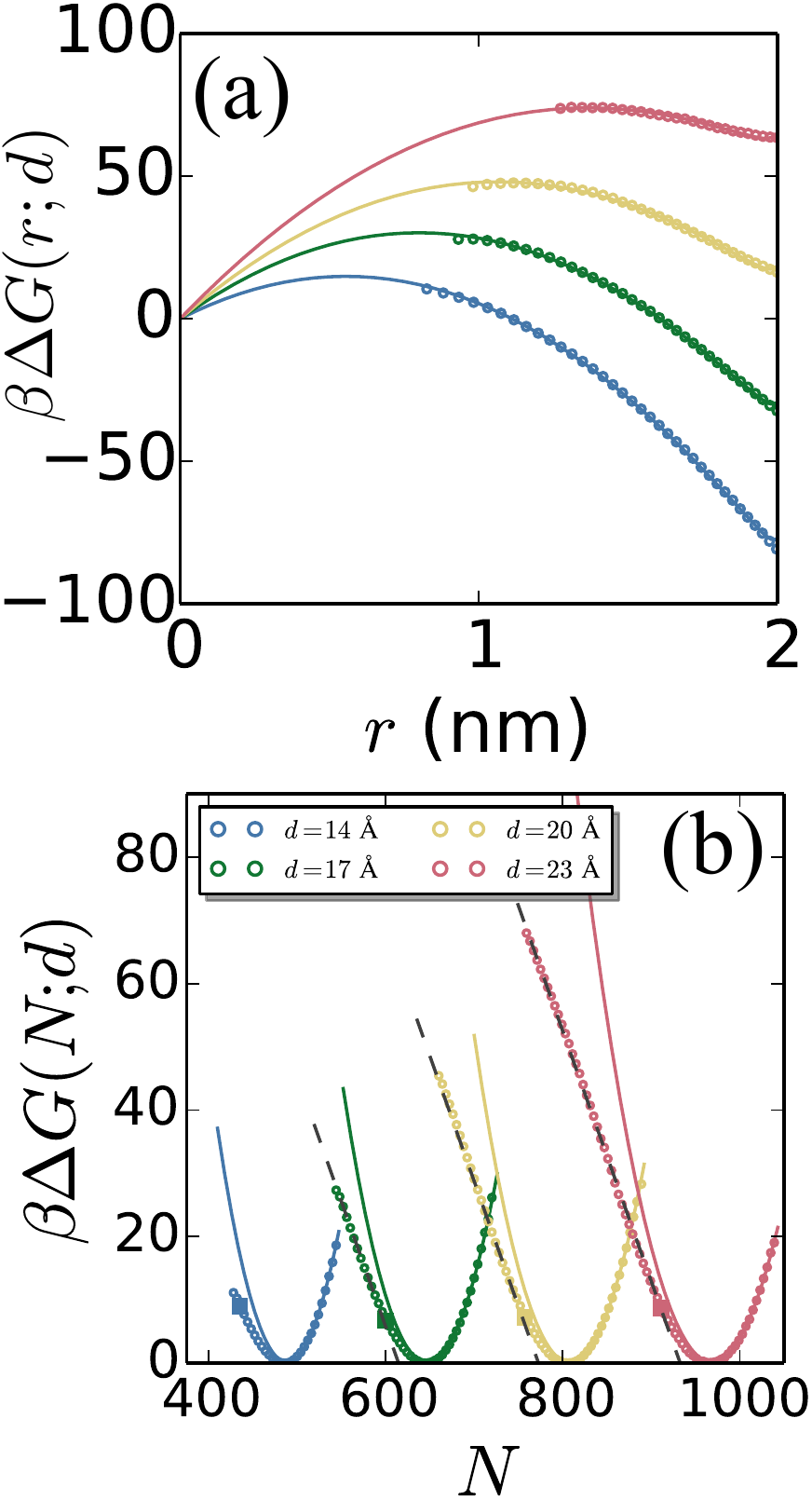}
\end{center}
\caption{
\label{fig:energetics} 
(a) $\beta\DG(N\le\Nkink;d)$ is recast as $\beta\DG(r;d)$, the free energy to form a vapor tube of radius, $r$. 
The points were obtained from the simulated free energy profiles by employing the relation $\pi r^2/L^2 = (\Nliq-N)/\Nliq$ in the region $r_{\rm kink}<r<L/2$, and the lines are fits to macroscopic theory. 
(b) The portion of the free energy corresponding to the liquid basin ($N\ge\Nkink$) is parabolic at high $N$ (Gaussian fluctuations) but linear at low $N$ (fat tails in water number distributions).
The crossover is gradual and occurs in the vicinity of $\Ncav$ (square symbols), that is, the value of $N$ for which $\langle \hcav\rangle_N$ is 0.5 with a negative slope.
The linear regions have roughly the same slope for all separations. 
%
%\vspace{-0.1in}
}
\end{figure}

%%%%%%%%%%%%%%%%%%%%%%%%%%%%%%%%%%%%
%
Given the abrupt change in the dewetted morphologies at $N=\Nkink$, we expect the functional form of the free energetics for $N>\Nkink$ and $N<\Nkink$ to be different.
%%%
Figures~\ref{fig:kink}c and ~\ref{fig:kink}d collectively show that configurations with $N\lesssim\Nkink$ feature a vapor tube spanning the confined region, consistent with classical arguments.
While the vapor tube undergoes extensive shape fluctuations, we find its average shape to be roughly cylindrical.
Coarse-grained density maps of select configurations reflecting the average vapor tube shape are included in the SI.
To compare the free energetics of the vapor tubes obtained from our simulations to macroscopic theory, we first transform the number of waters in the confined region, $N$, to an approximate vapor tube radius, $r$, using the simple relation, $\pi r^2/L^2 = (\Nliq-N)/\Nliq$.
The values of vapor tube radii thus obtained are consistent with the average radii of the vapor tubes observed in our simulations for $N<\Nkink$, as shown in the SI.
Having a one-to-one relation between $N$ and $r$ allows us to transform the simulated free energies, $\DG(N; d)$, into $r$-dependent free energies, $\DG(r;d)$ in the region $r_{\rm kink}<r<L/2$; the corresponding free energies are shown as symbols in Figure~\ref{fig:energetics}a.
The lines are fits to the $\DG(r)$-data using the macroscopic expectation, $\DG_{\rm fit}(r) = \Delta G_{\rm th}(r) + 2\kB T\ln(1-2r/L)$, where the logarithmic term corresponds to the translational entropy of the vapor tube.
$\DG(r)$ is fit separately for each $d$-value and yields values of $\gamma$ and $\lambda$ that are reasonable.
Values of $\gamma$ are in the range of $12.2 - 15.8\ \kB T/{\rm nm}^2$, comparable to the reported value of $14.5\ \kB T/{\rm nm}^2$ for the water model that we employ~\cite{Vega:JCP:2007}.
Our fits yield $-\lambda/\gamma$ in the $6.5 - 7.5\ \angstrom$ range, in accord with a recently reported experimental value of $\lambda=-30$~pN~\cite{Charlaix:PNAS:2012}, which yields $-\lambda/\gamma = 4.2\ \angstrom$.
These agreements are remarkable considering the simplicity of the model that we employ as well as the assumptions that we make (cylindrical vapor tube shape, constant surface tension independent of vapor tube curvature, etc.), suggesting that the energetics of the vapor tube are well-described by classical macroscopic theory.
Further details of our fitting procedure, the values of the fit parameters for each $d$, as well as our attempts to fit the simulation data to other reasonable expressions of $G_{\rm th}(r)$ can be found in the SI.
%%%

%%%
To investigate the free energetics of the isolated cavity ensemble, in Figure~\ref{fig:energetics}b, we focus on $\DG(N>\Nkink;d)$. 
In the liquid basin, that is, in the vicinity of $N=\Nliq$, the free energy (symbols) is parabolic (solid lines), indicating that the underlying density fluctuations are Gaussian.
While $\DG$ remains harmonic for $N>N_{\rm liq}$, it crosses over to being roughly linear (dashed lines) for $N < \Nliq$.
Such a crossover from parabolic to linear has also been observed adjacent to single extended hydrophobic surfaces~\cite{Patel:JPCB:2010,Rotenberg:JACS:2011,Patel:JPCB:2012}, and corresponds to the interfacial water undergoing a collective dewetting transition to expel water from a nanometer-sized cavity~\cite{LCW,LLCW,Patel:JPCB:2012}.
Indeed, as shown in Figure~\ref{fig:energetics}b, the location of the crossover agrees well with the corresponding $\Ncav$-values (squares) discussed in the previous section.
Interestingly, the slopes of the linear fat tails are similar for all $d$-values (in the range of $\Nkink+20\le N \le \Nliq-50$); the dashed lines shown in Figure~\ref{fig:energetics}b are linear fits with a slope of $ -0.396\ \kB T$ per water.
Additional details of the fitting procedure and the values of the parameters obtained are provided in the SI.
The difference between the values of $\Nliq$ and the $x$-intercepts of the linear fits is also approximately the same for all $d$-values, and is equal to $30\pm5$ waters.
Thus, the free energy for forming an isolated cavity of a given size (as quantified by the number of waters displaced from the confined region, $\Nliq-N$), is independent of the separation between the surfaces, that is, the free energetics of isolated cavity formation adjacent to one hydrophobic surface are largely unaffected by the presence of the other confining surface.
In contrast, the free energetics of vapor tube formation clearly depend on the inter-surface separation, $d$.
As a result, the location of the kink, where isolated cavities become metastable with respect to vapor tubes, also depends on $d$.
%
%%%

%%%%%%%%%%%%
%%%%%%%%%%%%

%%%%%%%%%%%%
\begin{figure*}
%\vspace{-0.1in}
\begin{center}
\includegraphics[width=0.98\textwidth]{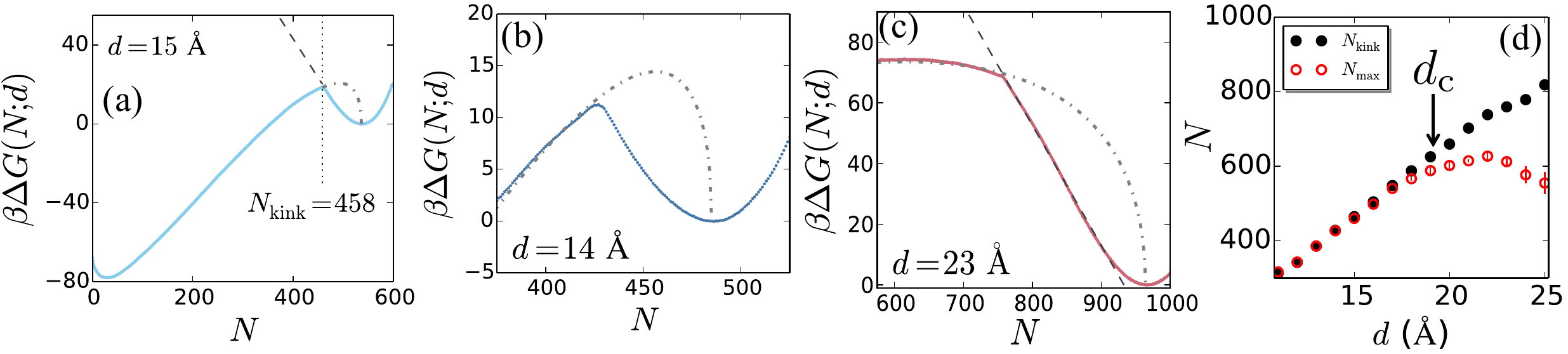}
\end{center}
%\vspace{-0.05in}
\caption{
\label{fig:mechanism} 
(a) The simulated $\beta\DG(N;d)$ for $d=15\angstrom$ (solid) is shown along with the expected metastable branches of the free energies corresponding to the vapor tube (dot-dashed) and isolated cavity (dashed) ensembles.
The metastable branches are anticipated from the fits shown in Figure~\ref{fig:energetics}.
The system minimizes its free energy by localizing to the ensemble with the lower free energy.
(b) For $d=14$~\AA, the nascent vapor tube formed at the kink is larger than the critical vapor tube anticipated by macroscopic theory, and therefore grows spontaneously.
As a result, the corresponding barrier to dewetting is smaller than that predicted by macroscopic theory.
(c) For larger separations, here $d=23$~\AA, the newly formed vapor tube is sub-critical, and has to grow larger for dewetting to proceed. 
(d) Comparison of the location of the kink, $\Nkink$, with the location of the barrier between the liquid and vapor basins, $N_{\rm max}$.
For small $d$, the barrier (point of highest $\DG$) occurs at the kink, so that $N_{\rm max}\approx\Nkink$.
In contrast, for larger $d$-values, the barrier occurs in the vapor tube segment of the simulated free energy profile and corresponds to the classical critical vapor tube, so that $N<\Nkink$.
The transition from non-classical to classical behavior occurs in the vicinity of the coexistence separation, $d_{\rm c}$.
%
%\vspace{-0.1in}
}
\end{figure*}
%%%%%%%%%%%%

%%%%%%%%%%%%%%%%%%%%%%%%%%%%%%%%%%%%
\section{Non-Classical Dewetting Mechanism Reduces Barriers} 
%%%%%%%%%%%%%%%%%%%%%%%%%%%%%%%%%%%%
%
Figure~\ref{fig:mechanism}a summarizes our findings for the dewetting mechanism presented thus far; in addition to the simulated $\DG(N;d)$ for $d=15\ \angstrom$, it highlights the metastable branches of the vapor tube and isolated cavity ensemble free energies, anticipated from the fits shown in Figure~\ref{fig:energetics}.
It is clear that the system minimizes its free energy at all times by staying on the branch with the lower free energy; at $N=\Nkink$, where the two free energy profiles intersect, the system jumps from the isolated cavity to the vapor tube ensemble.
Importantly, the non-classical path leading up to the formation of nascent vapor tubes ($N>\Nkink$) can result in smaller barriers to dewetting than anticipated by classical theory, as shown in Figure~\ref{fig:mechanism}b. 
For $d=14\angstrom$, the classical barrier (critical vapor tube) appears in the metastable segment of the free energy profile.
The system thus circumvents the classical barrier, and instead adopts the path involving isolated cavities, which give way to vapor tubes only at $N=\Nkink$; these nascent vapor tubes are larger than the critical vapor tube, so their subsequent growth is downhill in energy.
Thus, the barrier to dewetting is the free energetic cost for forming these nascent, super-critical vapor tubes, which is clearly smaller than the classical barrier.
In Figure~\ref{fig:mechanism}c, we illustrate that the nascent vapor tubes are not super-critical for all separations; for $d=23\angstrom$, the classical barrier appears in the stable segment of the free energy profile. 
Thus, while the kink in $\DG(N;d)$ again marks the formation of a vapor tube for $d=23\angstrom$, the vapor tube formed is smaller than the critical vapor tube, and must grow further in a process that is uphill in energy, before dewetting can proceed.
As a result, the barrier to dewetting coincides with that predicted by macroscopic theory.
%%%

%%%
To uncover the separation at which the system transitions from a supercritical to a subcritical nascent vapor tube, in Figure~\ref{fig:mechanism}d, we plot $\Nkink$ and $\Nmax$ as a function of $d$. 
Here, $\Nmax$ corresponds to the value of $N$ between the liquid and vapor basins where the $\DG(N;d)$ is the highest.
For small values of $d$, $\Nmax=\Nkink$, indicating that the $\DG(N;d)$ is the highest at the kink, consistent with the formation of supercritical nascent vapor tubes.
For larger values of $d$, there is an additional maximum in $\DG$ at $N_{\rm max}<N_{\rm kink}$, suggesting that the newly formed vapor tubes are smaller than the corresponding critical vapor tubes.
Interestingly, the separation at which the system transitions from non-classical to classical dewetting barriers is close to the separation at which there is coexistence between the liquid and vapor, $d_{\rm c}$. 
Thus, for the separations with a thermodynamically favorable driving force for dewetting, the mechanism for dewetting is manifestly non-classical, corresponding to the formation of super-critical vapor tubes from isolated cavities, and requiring a smaller free energetic barrier than anticipated by macroscopic theory.

\section{Discussion and Outlook} 
Our results highlight that water density fluctuations play a central role in the pathways to dewetting in hydrophobic confinement.
Enhanced water density fluctuations in the vicinity of hydrophobic surfaces stabilize isolated cavities relative to vapor tubes for $N>\Nkink$; the system resides in the classical vapor tube ensemble only for $N<\Nkink$.
While the free energetics of both the isolated cavity and vapor tube ensembles are well-described by the number of waters in confinement, $N$, this simple order parameter may not be sufficient to describe the transition from one ensemble to the other. 
Indeed, $\DG(N;d)$ represents a projection of a complex free energy landscape onto the single parameter, $N$, and a kink in $\DG(N;d)$ strongly suggests that the transition from an isolated cavity to a vapor tube involves order parameter(s) that are orthogonal to $N$.
While beyond the scope of the present work, it will be interesting to uncover these additional order parameters that define the transition state ensemble. 
It is conceivable that additional barriers may present themselves in the parameters that are orthogonal to $N$.
%%%

%%%
Dewetting in nanoscopic hydrophobic confinement plays an important role in biology; ranging from the assembly of multimeric proteins and the collapse of the hydrophobic protein core, to the vapor-lock gating of ion channels and the specific binding of ligands to hydrophobic grooves on their binding partners.
In particular, recent work has highlighted the importance of including the solvent coordinate, $N$, in describing the kinetics of hydrophobically-driven collapse and assembly~\cite{Mittal:JCP:2012,Li:JPCB:2012,Mondal:PNAS:2013,Setny:PNAS:2013}. 
Our results show that water density fluctuations stabilize non-classical pathways, which reduce the barriers along the $N$-coordinate, and should therefore enhance the kinetics of dewetting-mediated biophysical phenomena.
%%%

%%%
Dewetting in hydrophobic confinement is also important in a host of non-biological phenomena, ranging from heterogeneous nucleation of vapor bubbles and contact line pinning, to the Cassie-Wenzel transitions on textured surfaces~\cite{Kumar:JCP:2011}.
These phenomena involve intricate confinement geometries, which could result in complex pathways involving one or more transitions between various dewetted morphologies, such as the isolated cavities and vapor tubes that we observe here.
Fluctuation-mediated pathways ought to similarly reduce dewetting barriers associated with these diverse phenomena.
%
%% == end of paper:

%% Optional Materials and Methods Section
%% The Materials and Methods section header will be added automatically.
%% Enter any subheads and the Materials and Methods text below.
%
\section{Materials and Methods} 
We simulate SPC-E water in confinement between two square hydrophobic surfaces of size $L=4$~nm, for a range of separations, $d$ (See Figure~\ref{fig:model}a), ranging from $11\ \angstrom$ to $25\ \angstrom$, chosen to span the entire range of $d$-values with both liquid and vapor basins.
The surfaces are composed of 1008 atoms each, and are arranged on a hexagonal lattice with a spacing of $1.4\ \angstrom$.
The surface atoms interact with the water oxygens through the Lennard-Jones potential with the parameters, $\sigma=3.283\ \angstrom$ and $\epsilon=0.121$~kJ/mol; see refs.~\cite{Sharma:JPCB:2012,Sharma:PNAS:2012} for further details.
These interactions result in a water droplet contact angle $\theta\approx120^{\circ}$ on the hydrophobic surface, as shown in the SI.
We have chosen the SPC-E model of water~\cite{spce} since it adequately captures the experimentally known features of water such as surface tension, compressibility, and vapor-liquid equation of state near ambient conditions, which are important in the study of hydrophobic effects~\cite{Varilly:JPCB:2013,Chandler:Nature:2005}.  
Our simulations contain roughly 10,000 to 15,000 water molecules and were performed in the NVT ensemble, thermostatted at $T=300$~K using the canonical velocity rescaling thermostat~\cite{Bussi:JCP:2007}.
We employ a periodic simulation box with the hydrophobic surfaces of interest fixed at the center of the box, and a buffering liquid-vapor interface nucleated at the top of the box with the help of a wall of purely repulsive particles.
The buffering interface ensures that the system is at the saturation pressure of SPC-E water at 300K~\cite{Miller:PNAS:2007,Patel:JPCB:2010}; free energies obtained with such a construct have been shown to be nearly indistinguishable from those obtained in the NPT ensemble at a pressure of $1\,$bar~\cite{Patel:JSP:2011}.  
Short-ranged interactions were truncated at 1~nm, whereas long-ranged electrostatic interactions were computed using the particle mesh Ewald method~\cite{PME}.
The bonds in water were constrained using SHAKE~\cite{SHAKE}. 
To study the free energetics of dewetting, we select the cuboid shaped ($L \times L \times d$) observation volume between the hydrophobic surfaces, and estimate the free energies, $\Delta G(N;d)$,  using the indirect umbrella sampling (INDUS) method~\cite{Patel:JPCB:2010,Patel:JSP:2011}.
Each biased simulation was run for 6~ns and the first 1~ns was discarded for equilibration.
%
%\vspace{-0.1in}
%\end{materials}
%
\begin{acknowledgments}
RCR~and~AJP were supported in part by financial support from the National Science Foundation (NSF) through a Seed Grant from the University of Pennsylvania Materials Research Science and Engineering Center (NSF UPENN MRSEC DMR 11-20901).
SG was supported by the NSF (CBET-1159990).  
PGD gratefully acknowledges support from the NSF (CBET-1263565 and CHE-1213343).
\end{acknowledgments}
%

%% PNAS does not support submission of supporting .tex files such as BibTeX.
%% Instead all references must be included in the article .tex document. 
%% If you currently use BibTeX, your bibliography is formed because the 
%% command \verb+\bibliography{}+ brings the <filename>.bbl file into your
%% .tex document. To conform to PNAS requirements, copy the reference listings
%% from your .bbl file and add them to the article .tex file, using the
%% bibliography environment described above.  

%% Enter the largest bibliography number in the facing curly brackets
%% following \begin{thebibliography}

%\bibliography{2step}
%

%\end{article}

%% file: SI-Feb16-2015-submit.tex
\setcounter{figure}{0}
\setcounter{section}{0}

\title{Supplementary Information for ``Pathways to dewetting in hydrophobic confinement''}

%Author List
\author{Richard C. Remsing}
\altaffiliation[]{RCR and EX are joint first authors who contributed equally to this work}
\affiliation{Department of Chemical \& Biomolecular Engineering, University of Pennsylvania, Philadelphia, PA 19104, USA}

\author{Erte Xi}
\altaffiliation[]{RCR and EX are joint first authors who contributed equally to this work}
\affiliation{Department of Chemical \& Biomolecular Engineering, University of Pennsylvania, Philadelphia, PA 19104, USA}

\author{Srivathsan Vembanur}
\affiliation{Howard P. Isermann Department of Chemical \& Biological Engineering, and Center for Biotechnology and Interdisciplinary Studies, Rensselaer Polytechnic Institute, Troy, NY 12180, USA}

\author{Sumit Sharma}

\author{Pablo G. Debenedetti}
\affiliation{Department of Chemical \& Biological Engineering, Princeton University, Princeton, NJ 08544, USA}

\author{Shekhar Garde}
\affiliation{Howard P. Isermann Department of Chemical \& Biological Engineering, and Center for Biotechnology and Interdisciplinary Studies, Rensselaer Polytechnic Institute, Troy, NY 12180, USA}

\author{Amish J. Patel}
\email{amish.patel@seas.upenn.edu}
\affiliation{Department of Chemical \& Biomolecular Engineering, University of Pennsylvania, Philadelphia, PA 19104, USA}

% Non-PNAS format
%\author{Richard C. Remsing}
%\author{Erte Xi}
%\affiliation{Department of Chemical \& Biomolecular Engineering, University of Pennsylvania, Philadelphia, PA 19104, USA}
%\author{Srivathsan Ranganathan}
%\affiliation{Howard P. Isermann Department of Chemical \& Biological Engineering, and Center for Biotechnology and Interdisciplinary Studies, Rensselaer Polytechnic Institute, Troy, NY, 12180, USA}
%\author{Sumit Sharma}
%\author{Pablo Debenedetti}
%\affiliation{Department of Chemical Engineering, Princeton University, Princeton, NJ 08544, USA}
%\author{Shekhar Garde}
%\affiliation{Howard P. Isermann Department of Chemical \& Biological Engineering, and Center for Biotechnology and Interdisciplinary Studies, Rensselaer Polytechnic Institute, Troy, NY, 12180, USA}
%\author{Amish J. Patel}
%\affiliation{Department of Chemical \& Biomolecular Engineering, University of Pennsylvania, Philadelphia, PA 19104, USA}

%\date{\today}
\maketitle

%\begin{article}

%****************************************** New Section *********************************************************%

\section{Estimating ``Smoothed'' Derivatives of the Free Energy}

\begin{figure}[tb]
\begin{center}
\includegraphics[width=0.49\textwidth]{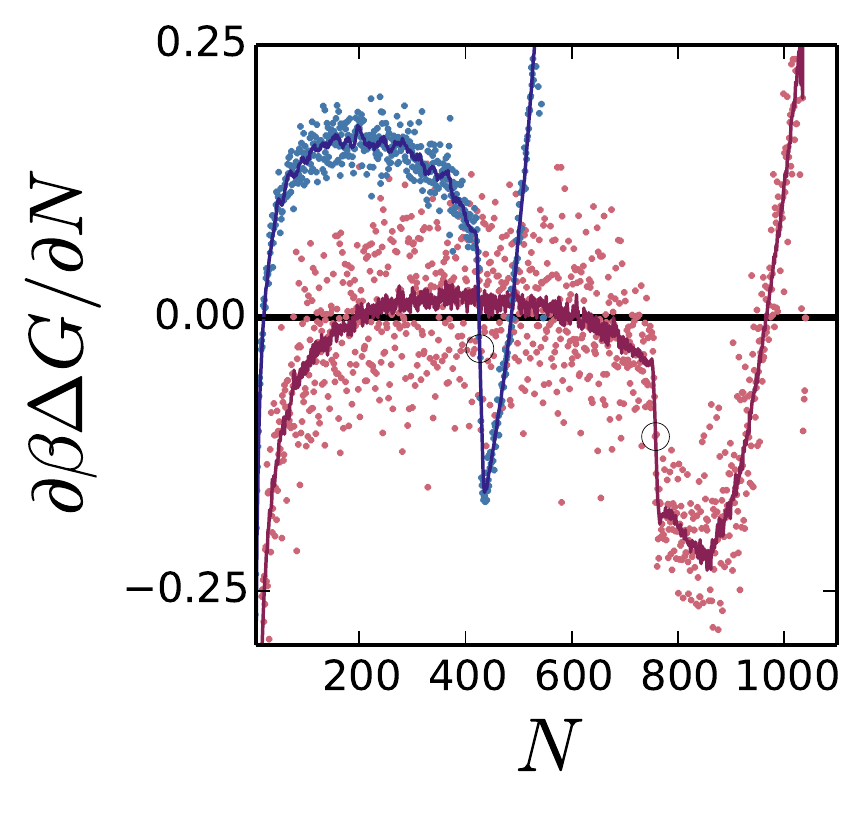}
\end{center}
\caption[Free Energy Derivatives]
{
Derivatives of the unsmoothed (points) and smoothed (lines) free energies for $d=14$~\AA \ (blue) and $d=23$~\AA \ (red).
}
\label{fig:dfe}
\end{figure}

To demonstrate the existence of kinks in $\Delta G(N; d)$ more clearly, we plot smoothed derivatives of $\Delta G(N; d)$ in Figure~2b of the main text.
%; the derivatives display a sharp (nearly discontinuous) change in slope.
%In order to calculate derivatives of the free energy, $\partial \beta\Delta G/\partial N$, as shown in Figure~2 of the main text, we smooth the simulation data.
%
Directly computing derivatives of $\Delta G(N)$ using finite difference, for example, using $\partial \Delta G/\partial N = \Delta G(N+1)-\Delta G(N)$, results in noisy estimates due to the numerical error in $\Delta G(N)$ (Figure~\ref{fig:dfe}).
To minimize such effects, we first smooth $\Delta G(N)$ using a rectangular window function, such that
\begin{equation}
\Delta G_{\rm smoothed}(N) = \frac{1}{\Delta N}\sum_{N-\Delta N/2}^{N+\Delta N/2 -1} \Delta G(N)
\end{equation}
is the smoothed free energy.
Smoothed derivatives are then estimated from finite differences of such smoothed free energies,
\begin{equation}
\frac{\partial \Delta G}{\partial N}(N) \approx \brac{\Delta G_{\rm smoothed}(N+1)-\Delta G_{\rm smoothed}(N)}.
\end{equation}
Derivatives of both the unsmoothed and the smoothed (using a width of $\Delta N=10$) free energies for select values of $d$ are compared in Figure~\ref{fig:dfe}.

\begin{figure}[tb]
\begin{center}
\includegraphics[width=0.49\textwidth]{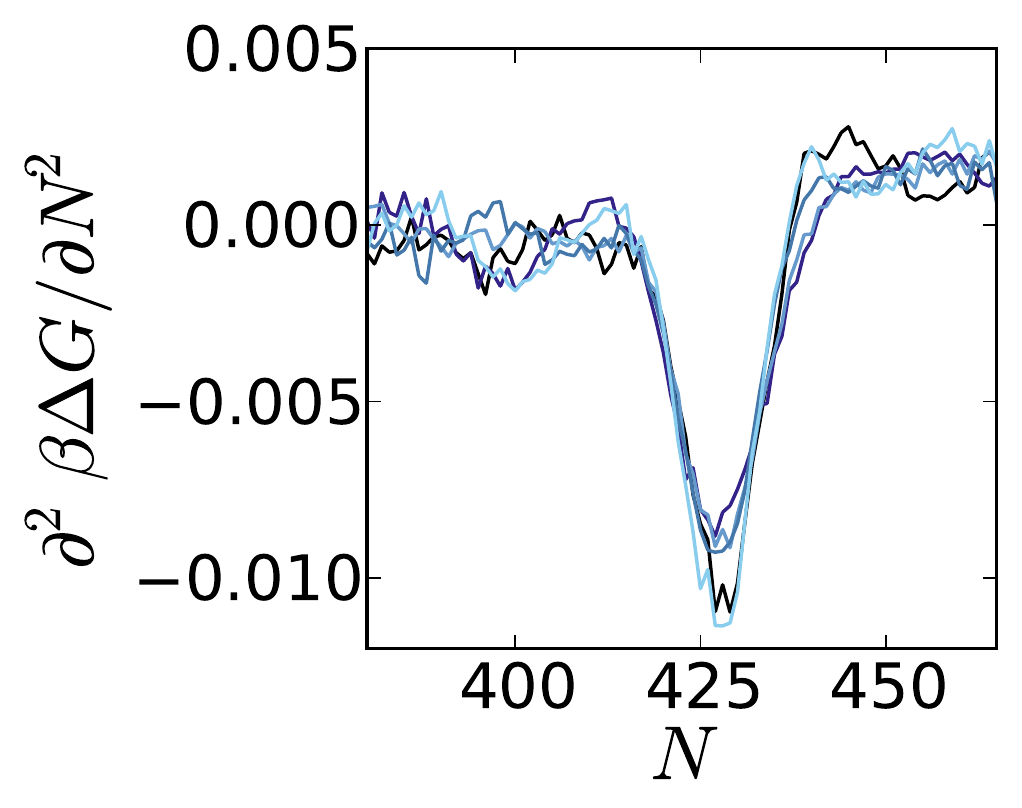}
\end{center}
\caption[Second Derivate]
{
Second derivative of the free energy for $d=14$~\AA. Block averaging was used to obtain $N_{\rm kink}$ and its associated error bars; the derivatives of each of the the five blocks are shown here.
$N_{\rm kink}$ is the location of the minimum in $\partial^2 \beta\Delta G/\partial N^2$.
}
\label{fig:d2}
\end{figure}

An analogous smoothing of the first derivative was also performed in order to obtain the second derivative of the free energy, which in turn enables us to  robustly estimate $N_{\rm kink}$ as the location of the minimum in $\partial^2 \beta\Delta G/\partial N^2$.
The error in $N_{\rm kink}$ was estimated using block averaging; the trajectories were divided into five blocks of 1 ns each, and $N_{\rm kink}$ was obtained for each of the five blocks. 
As an example, the second derivatives for five such blocks are shown in Figure~\ref{fig:d2} for $d=14$~\AA.

%****************************************** New Section *********************************************************%
\section{Obtaining Instantaneous Interfaces that Envelop Dewetted Regions}
Here, we closely follow and build upon the approach for estimating instantaneous interfaces, originally developed by Willard and Chandler~\cite{Willard:JPCB:2010}.
In addition to considering the coarse-grained water density, we also include a coarse-grained density arising from the plate atoms into an overall, normalized coarse-grained density at $(x,y,z)$ as follows:
\begin{equation}
\tilde{\rho}_{\rm total}(x,y,z;\Rbar) = \frac{\tilde{\rho}_{\rm water}(x,y,z;\Rbar)}{\rho^{\rm bulk}_{\rm water}} +  \frac{\tilde{\rho}_{\rm plate}(x,y,z;\Rbar)}{\rho^{\rm max}_{\rm plate}}.
\label{eq:ii1}
\end{equation}
Here, $\Rbar$ represents the positions of all the heavy atoms in the system in a given configuration, the coarse-grained water density is normalized by the corresponding bulk density, and the coarse-grained plate density by the maximum coarse-grained plate density that occurs at the center of the plate.
%
%The sum of these two normalized densities fields give us the coarse-grained density of the whole system,
%
Defined this way, $\tilde{\rho}_{\rm total}$ is approximately equal to unity everywhere in the liquid state and near unity at the center of the plates and in the interfacial region.
Configurations containing dewetted regions (cavities) will have significantly smaller values of $\tilde{\rho}_{\rm total}$ that approach zero in the vicinity of the cavity.
Therefore, the $\tilde{\rho}_{\rm total}=0.5$ iso-density surface serves as a convenient definition of the instantaneous interface, allowing us to readily visualize the position, size, and shape of the cavity;
we use the Marching Cube algorithm~\cite{marching_cubes} to identify the instantaneous interface.
The coarse-grained density fields of the individual species are estimated using
\begin{equation}
\tilde{\rho}_{\alpha}(x,y,z;\Rbar) = \sum_{i=1}^{N_{\alpha}} \phi(x_i-x)\phi(y_i-y)\phi(z_i-z),
\label{eq:ii2}
\end{equation}
where $\alpha$ represents either the water oxygen atoms or the plate atoms, $N_{\alpha}$ is the number of atoms of type $\alpha$, and $(x_i,y_i,z_i)$ correspond to the coordinates of atom $i$.
%%%

%%%
For each configuration obtained from our simulations, we set up a three-dimensional grid to compute the coarse-grained density field with 0.1~nm spacing in each dimension. 
The coarse-graining function $\phi(x)$ is chosen to be a Gaussian with a width of $0.24$~nm, which was truncated at $0.7$~nm, shifted down to make it continuous, and normalized.
The particular characteristics of the instantaneous interfaces thus computed, as well as those of the dewetted regions, such as their exact shapes and volumes, will depend on the choices made in Equations~\ref{eq:ii1} and~\ref{eq:ii2}, as well as the parameters chosen.
A discussion of how these choices affect the instantaneous interface calculation and which choices are judicious is beyond the scope of this work, and will be the subject of a separate publication.
However, it is important to note that the qualitative insights that we obtain in this work are not sensitive to the particular choices that we make here.

%****************************************** New Section *********************************************************%
\section{Movies}

\begin{figure}
\includegraphics[width=0.49\textwidth]{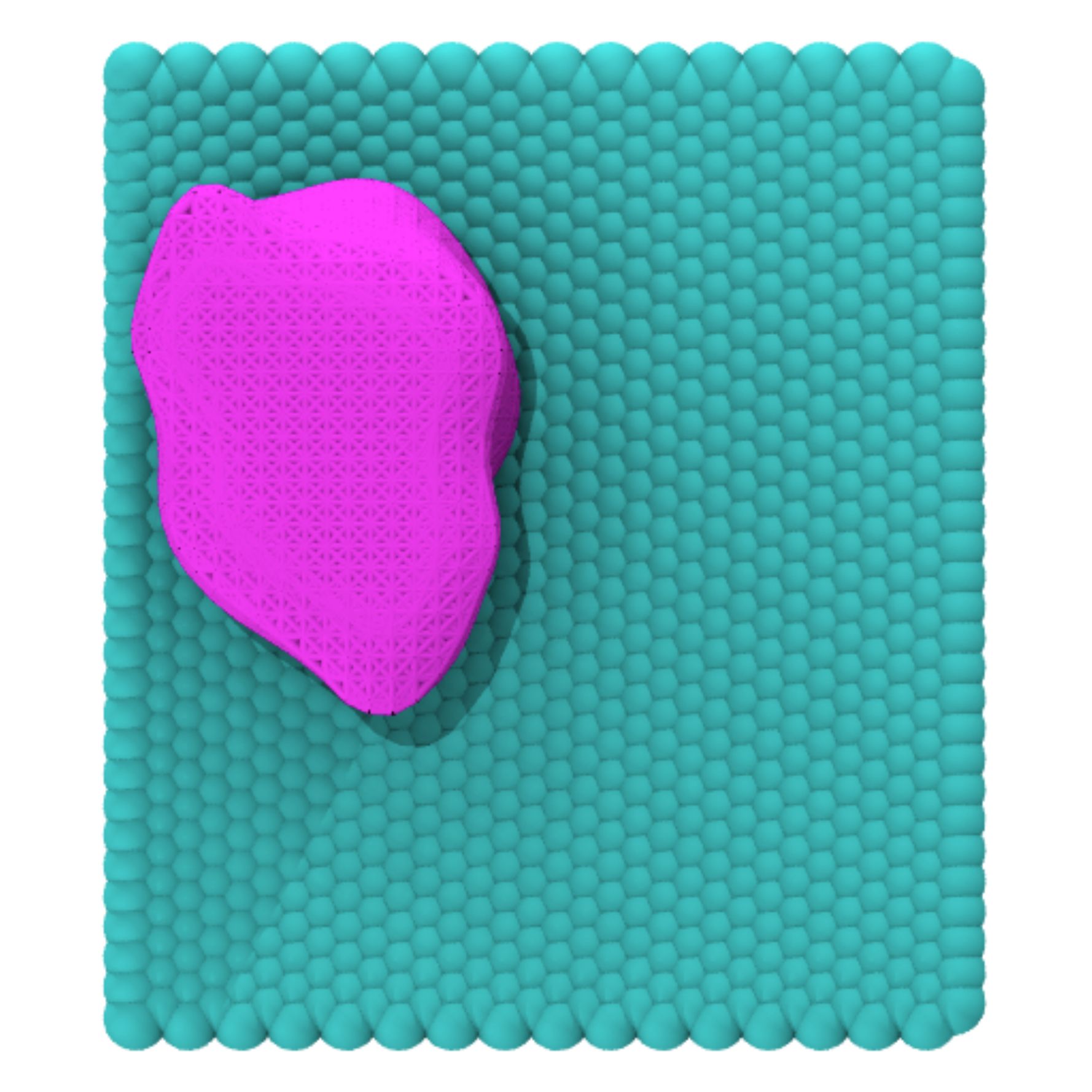}
\includegraphics[width=0.49\textwidth]{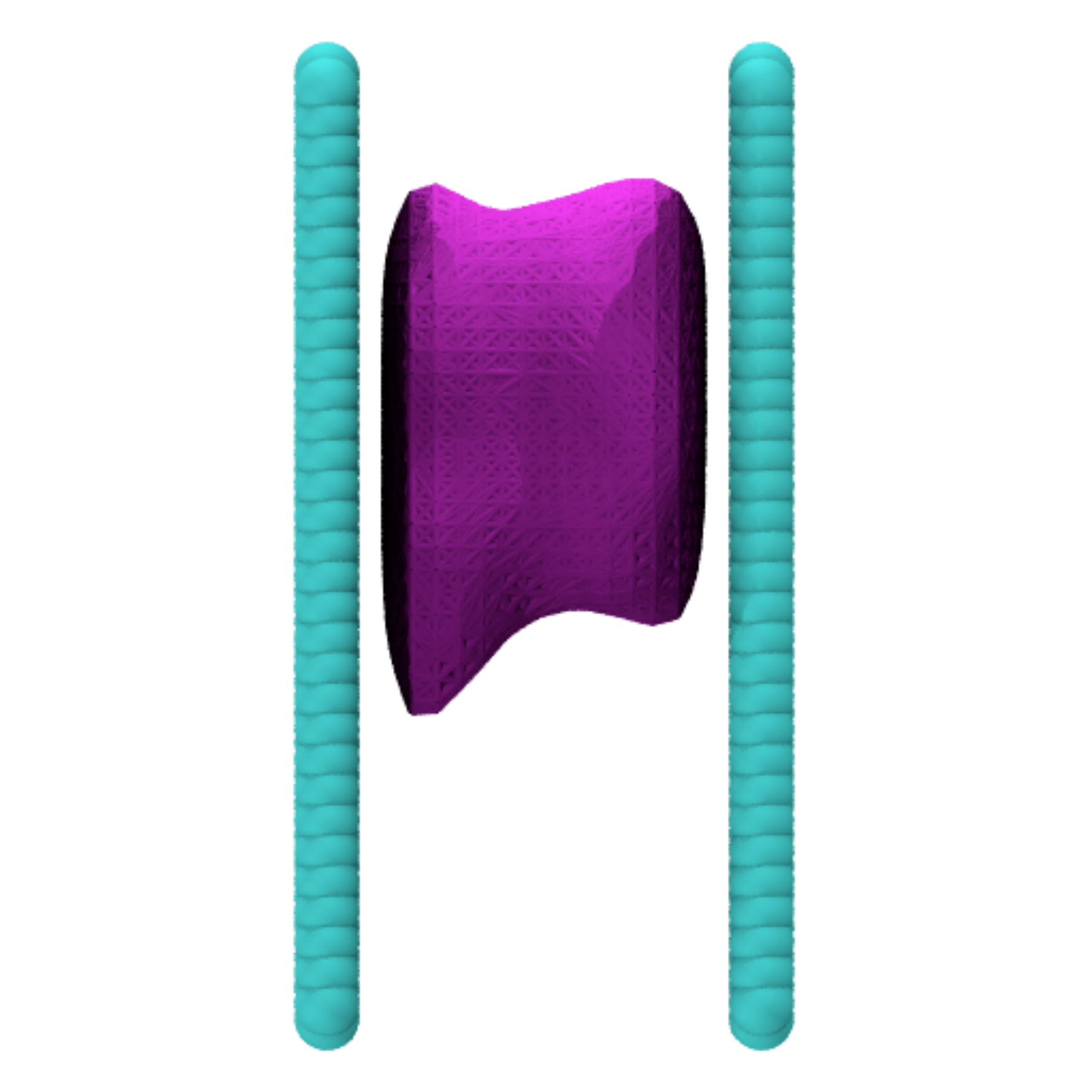}
%\includemovie[poster=N650_front.pdf,repeat]{6cm}{6cm}{d20II_N650_front_hd.mpg}
%\includemovie[poster=N650_side.pdf,repeat]{6cm}{6cm}{d20II_N650_side_hd.mpg}
\caption{
Movies (click to view) illustrating a plate-spanning vapor tube in a biased simulation with $\tilde{N}^*=650$ and $\kappa=0.12$~kJ/mol,
with $\avg{\tilde{N}} = 653$.
Plate atoms are shown as spheres, while the purple mesh corresponds to the instantaneous interface enveloping the vapor region
(water molecules have been omitted for clarity). The movie is shown from directions (top) perpendicular and (bottom) parallel to the plane of the plates.
The duration of the movie corresponds to 250~ps of simulation time. Adobe Reader may be needed to view.
Representative snapshots are shown here; movie files are available upon request and will appear in the final published version.
}
\label{fig:movie650}
\end{figure}

The dynamical nature of the vapor tubes described in the main text can be observed in the movie shown in Figure~\ref{fig:movie650}.
There, we show a top and a side view of a plate-spanning vapor tube from a simulation with a biasing potential, $\kappa(\tilde{N} - \tilde{N}^*)^2 / 2$; $\kappa=0.12$~kJ/mol and $\tilde{N}^*=650$. 
The average value of $\tilde{N}$ in the presence of the biasing potential was $\avg{\tilde{N}} = 653$.
The plate atoms are shown as cyan spheres, water molecules are not shown for clarity, and the purple mesh corresponds to the instantaneous interface enveloping the vapor region.
In essence, the purple mesh defines regions devoid of water molecules, such that the space outside the mesh is high in water density.

\begin{figure}
\includegraphics[width=0.49\textwidth]{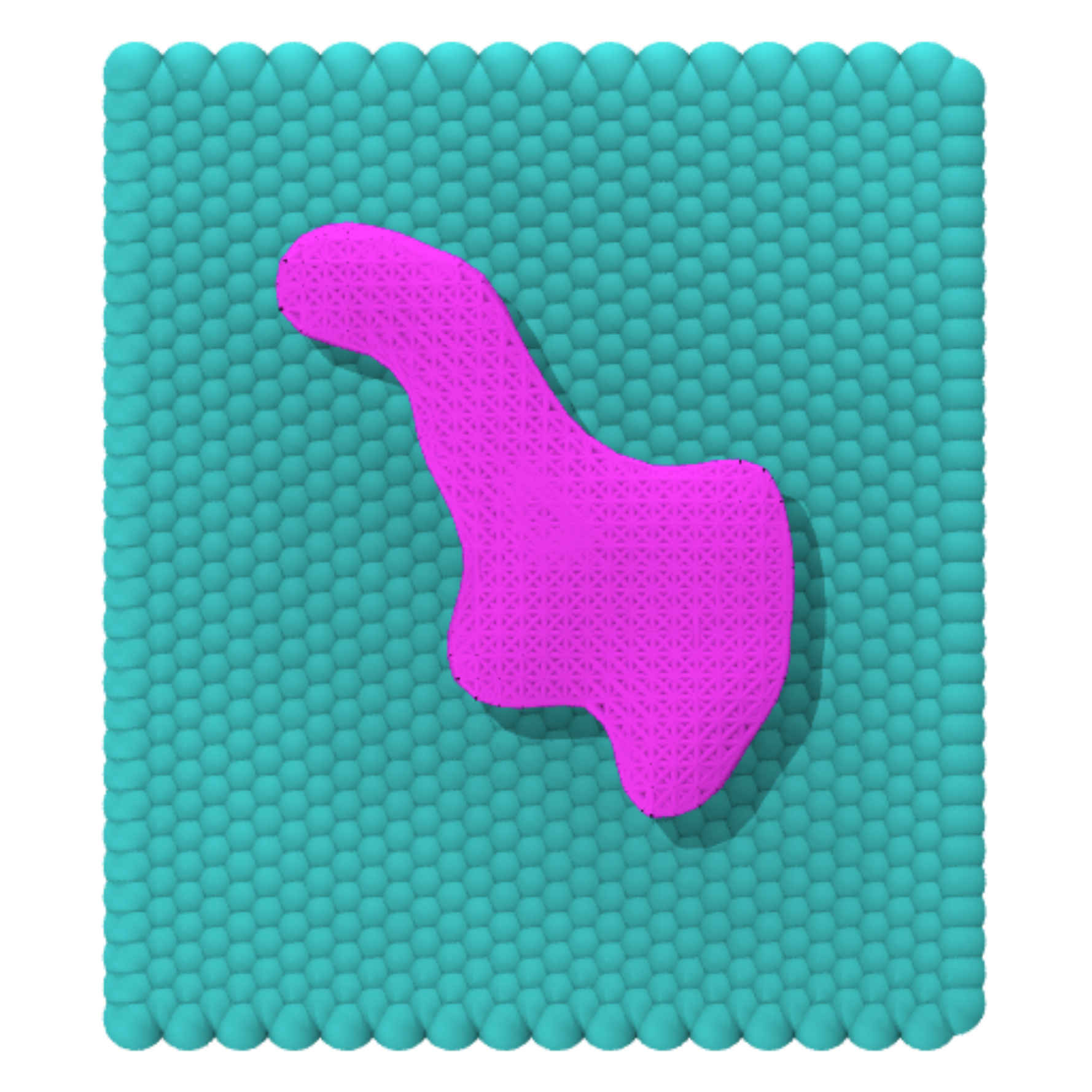}
\includegraphics[width=0.49\textwidth]{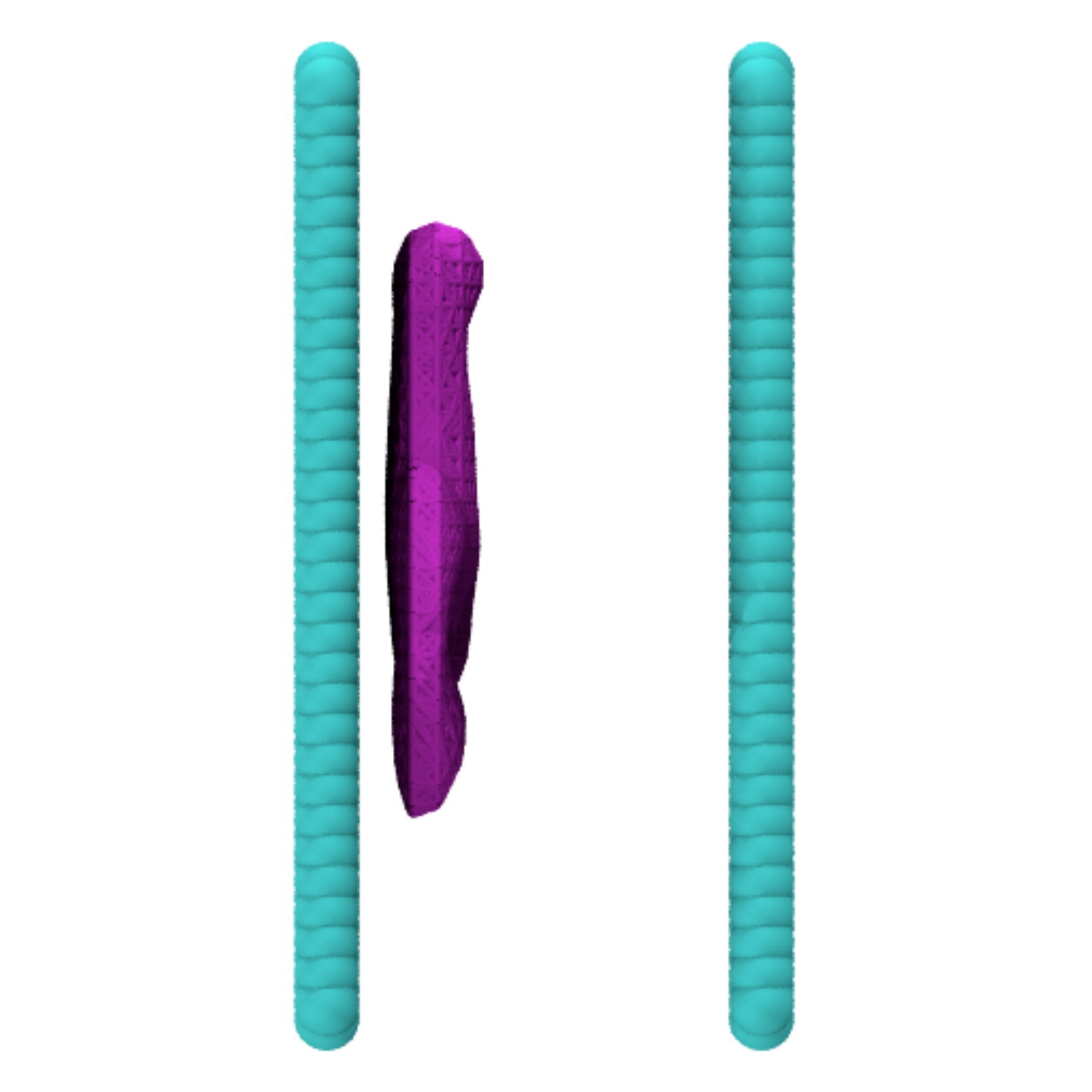}
%\includemovie[poster=N660_front.pdf,repeat]{6cm}{6cm}{d20II_N660_front_hd.mpg}
%\includemovie[poster=N660_side.pdf,repeat]{6cm}{6cm}{d20II_N660_side_hd.mpg}
\caption{
Movies (click to view) illustrating vapor bubbles formed at the surface of a hydrophobic plate in a biased simulation with $\tilde{N}^*=660$ and $\kappa=0.12$~kJ/mol,
with $\avg{\tilde{N}}=668$.
Plate atoms are shown as spheres, while the purple mesh corresponds to the instantaneous interface enveloping the vapor region
(water molecules have been omitted for clarity). The movie is shown from directions (top) perpendicular and (bottom) parallel to the plane of the plates.
The duration of the movie corresponds to 250~ps of simulation time. Adobe Reader may be needed to view.
Representative snapshots are shown here; movie files are available upon request and will appear in the final published version.
}
\label{fig:movie660}
\end{figure}

We show similar movies for an isolated cavity at the surface(s) of the hydrophobic plate(s) in Figure~\ref{fig:movie660}.
The coloring scheme is the same as in Figure~\ref{fig:movie650},
and the configurations are taken from a biased simulation with $\tilde{N}^*=660$ and $\kappa=0.12$~kJ/mol; the corresponding $\avg{\tilde{N}}=668$.
There, an isolated cavity can be observed fluctuating in shape and size, and even moving from one plate to the other.
The mechanism for bubble migration from one plate to another is an open question relevant to dewetting transitions and deserves further investigation.

%****************************************** New Section *********************************************************%
\section{Definition of the Tube and Cavity Indicator Functions}
The tube indicator function $h_{\rm tube}$ is determined by examining the coarse-grained density field, $\tilde{\rho}_{\rm total}$, between the two plates.
We define the $x$-coordinate to be perpendicular to the plates, such that the two plates are located at $x_1$ and  $x_2$, respectively, with $x_1<x_2$.
We additionally define a buffer region $b=0.4$~nm from the center of each plate to avoid the region where the coarse-grained density originating
from the plate atoms is larger than or close to 0.5.
If at some location, ($y^*, z^*$), the total coarse-grained density is below 0.5 at all $x$-values between $x_1+b$ and $x_2-b$, we assign $h_{\rm tube}=1$, \ie
\begin{align}
h_{\rm tube} &= 1 \ {\rm if} \  \exists \ (y^*,z^*) \nonumber \\
& \ \ \  : \curly{ \tilde{\rho}_{\rm total}(x,y^*,z^*)<0.5} \ \forall \ x\in (x_1+b,x_2-b) \nonumber \\
&= 0 \ \ {\rm otherwise}.
\end{align}

The cavity indicator function $h_{\rm cav}$ is similarly determined from $\tilde{\rho}_{\rm total}$.
However, $h_{\rm cav}$ is equal to unity if the coarse-grained density is less than a half at some location between the plates, but not for all $x$-values between the plates.
In other words,
\begin{align}
%h_{\rm cav} &= 1 \ {\rm if} \  \exists \ (x^*,y^*,z^*) \ : \curly{ \tilde{\rho}_{\rm total}(x^*,y^*,z^*)<0.5}, \ \exists \ x \ : \curly{ \tilde{\rho}_{\rm total}(x,y^*,z^*)>0.5}  \\
h_{\rm cav} &= 1 \ {\rm if} \  \exists \ (x^*,y^*,z^*) \nonumber \\
& \ \ \ : \curly{ \tilde{\rho}_{\rm total}(x^*,y^*,z^*)<0.5}, \ h_{\rm tube}=0  \nonumber \\
&= 0 \ \ {\rm otherwise}.
\end{align}

%****************************************** New Section *********************************************************%
\section{Procedure for Calculating Unbiased Ensemble Averages}

All umbrella sampling simulations performed in this work employ a biasing potential.
% the coarse-grained variable $\tilde{N}$.
Therefore, when performing ensemble averages of any observable, care must be exercised in accounting for the bias introduced by the potential.
This is done using the WHAM/MBAR formalism~\cite{WHAM,UWHAM,MBAR},
such that the unbiased ensemble average of any observable, $A(\Rbar)$, which can be expressed as a function of the configuration vector, $\Rbar$, is given by
\begin{equation}
\avg{A(\Rbar)} = C^{-1}\sum_{j=1}^K \sum_{n=1}^{N_j} \frac{A(\Rbar_{j,n})}{\sum_{k=1}^K N_k e^{\beta F_k - \beta V_k(\Rbar_{j,n})}},
\label{eq:mbaravg}
\end{equation}
where
\begin{equation}
C = \sum_{j=1}^K \sum_{n=1}^{N_j} \frac{1}{\sum_{k=1}^K N_k e^{\beta F_k - \beta V_k(\Rbar_{j,n})}}.
\end{equation}
Here, $ F_k$ is the free energy of the $k$th biased ensemble,
$K$ is the number of biasing potentials or windows used, $N_k$ is the number of samples in window $k$, $V_k$ is the biasing potential for window $k$,
and $\Rbar_{j,n}$ refers to configuration $n$ in window $j$.

Equation~\ref{eq:mbaravg} is used here to calculate the ensemble average of $h_{\rm tube}$, conditioned on the number of waters in confinement being $N$, according to
%\begin{align}
\begin{equation}
\avg{h_{\rm tube}(\Rbar)}_N 
%&= 
=
\frac{
\avg{h_{\rm tube}(\Rbar)\delta_{N,N(\Rbar)}} 
}{ 
\avg{\delta_{N,N(\Rbar)}}
}
%\\
%&= 
%C^{-1}\sum_{j=1}^K \sum_{n=1}^{N_j} \frac{h_{\rm tube}(\Rbar_{j,n}) \delta_{N,N(\Rbar_{j,n})}}{\sum_{k=1}^K N_k e^{\beta F_k - \beta V_k(\Rbar_{j,n})}},
\label{eq:whamtube}
\end{equation}
%\end{align}
where $\delta_{N,N(\Rbar_{j,n})}$ is the Kronecker delta function.
The ensemble average of $h_{\rm cav}$ as a function of $N$ is calculated in an analogous manner by replacing $h_{\rm tube}(\Rbar)$ with $h_{\rm cav}(\Rbar)$ in Equation~\ref{eq:whamtube}.

%****************************************** New Section *********************************************************%
\section{Average Shape of the Vapor Tube}
\begin{figure}[tb]
\begin{center}
\includegraphics[width=0.49\textwidth]{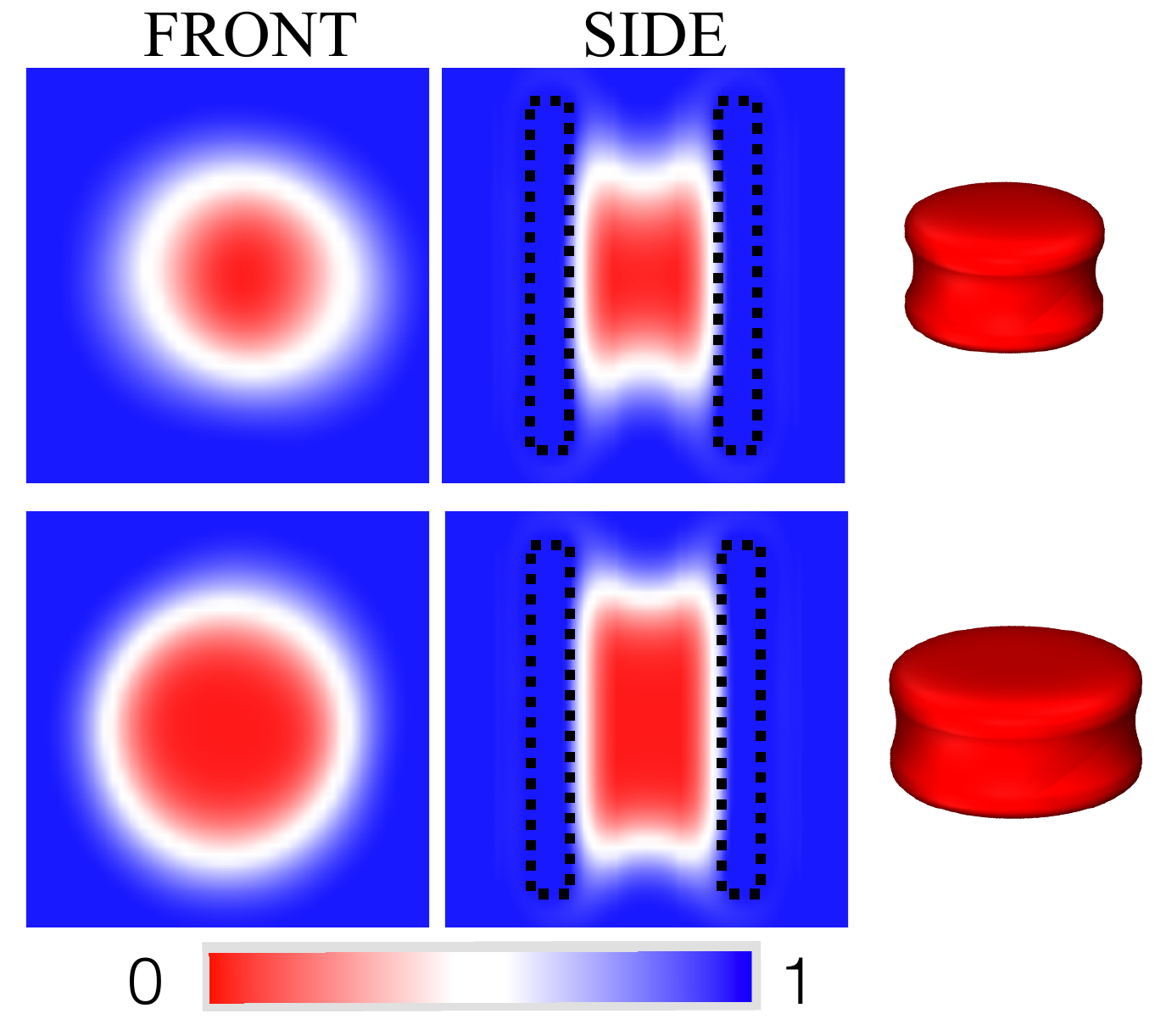}
\end{center}
\caption[Vapor Tube Shape]
{
The average coarse-grained density, $\tilde{\rho}_{\rm tot}(x,y,z)$, for $d=20$~\AA, with an average of 569 waters (top) and 474 waters (bottom) between the plates.
The top panel was obtained from a biased simulation with $\tilde{N}^*=570$ and $\kappa=0.03$~kJ/mol, and the bottom panel was obtained from a biased simulation
with $\tilde{N}^*=480$ and $\kappa=0.03$~kJ/mol
The rightmost panels depict three-dimensional renderings of the vapor tube shape at the iso-density surface $\tilde\rho_{\rm tot} = 0.5$
}
\label{fig:shape}
\end{figure}

The average shape of a vapor tube is identified by first taking the average of the coarse grained density, $\tilde{\rho}_{\rm tot}(x,y,z)\equiv\avg{\tilde{\rho}_{\rm total}(x,y,z;\Rbar)}$, over 5000 frames.
Figure~\ref{fig:shape} displays the averaged density of the $d=20$~\AA \, with an average of 569 waters (top) and 474 waters (bottom) between the plates,
obtained from biased simulations with $\tilde{N}^*=570$ and $\kappa=0.03$~kJ/mol and $\tilde{N}^*=480$ and $\kappa=0.03$~kJ/mol, respectively.
From left to right in each row presents a front view, side view and three-dimensional view of the averaged shape of the vapor tube.
The three-dimensional rendering of the vapor tube is obtained as the iso-density surface $\tilde\rho_{\rm tot} = 0.5$, and this surface provides an accurate description of the average shape of the vapor tube.
Additionally, the radii obtained directly from these isodensity surfaces, $1.03$~nm and $1.30$~nm, respectively for the top and bottom panels of Figure~\ref{fig:shape},
are in reasonably good agreement with those obtained from the simple relation between $r$ and $N$ used in the main text (Equation~\ref{eq:radN}), $1.22$~nm and $1.45$~nm, respectively.
Note that the vapor tubes are cylindrical to a good approximation, in accord with macroscopic theory.
However, further corrections to such theories could be obtained by taking into account finer details of the ``hour-glass'' shape of the vapor tubes.
Because the precise shape of the vapor tubes depends on the details of the instantaneous interface calculation and the parameters employed, we do not attempt such corrections here.

%****************************************** New Section *********************************************************%
\section{Contact Angle Determination}

The contact angle of the surface was determined by performing a 2~ns simulation of a cylindrical droplet containing 4142 water molecules.
This cylindrical droplet was divided into five slabs, each 1~nm in width, and the density as a function of the radius $y$ and the height $z$ of each slab was computed. 
The droplet profile is then defined as the point where the density of the droplet is equal to half that of the bulk density. 

\begin{figure}[tb]
\begin{center}
\includegraphics[width=0.4\textwidth]{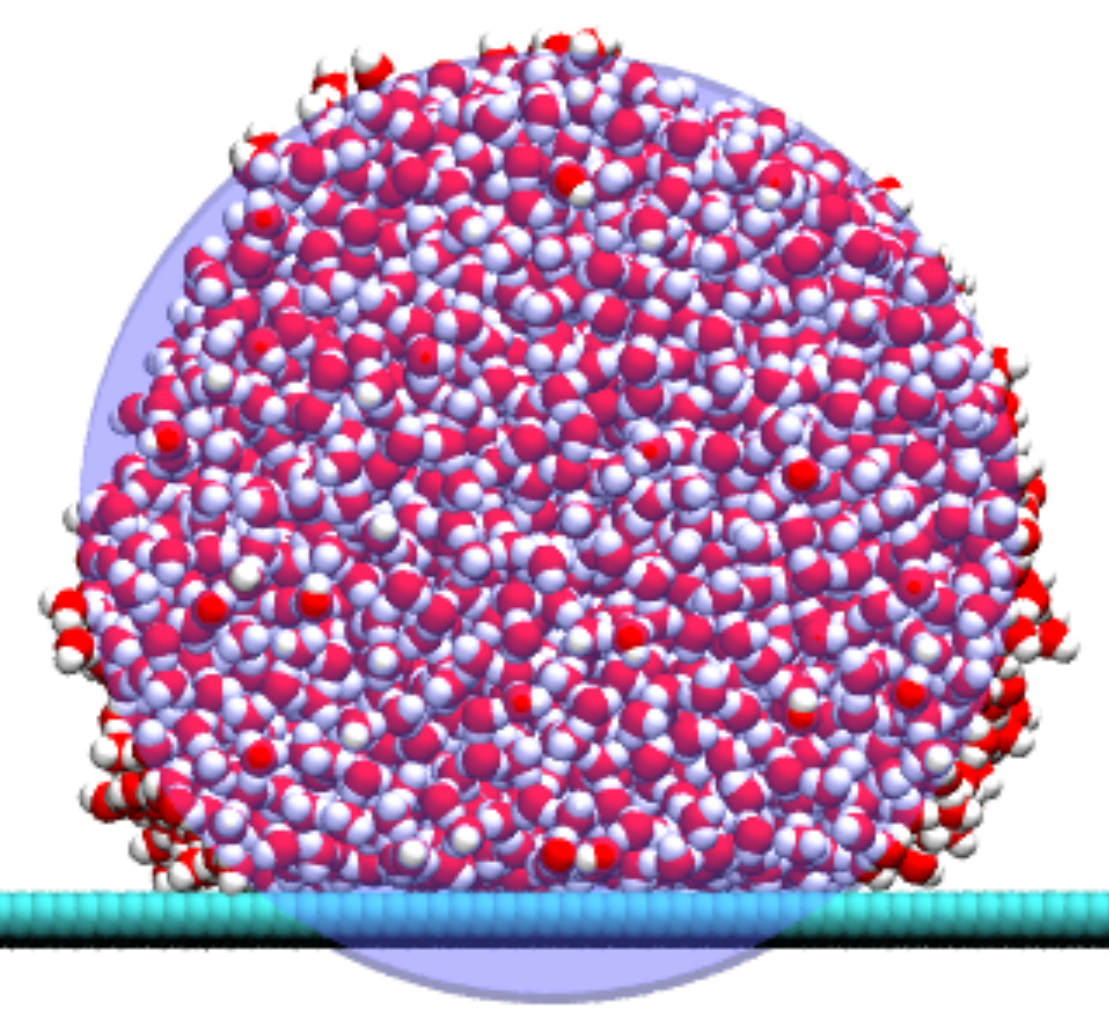}
\includegraphics[width=0.49\textwidth]{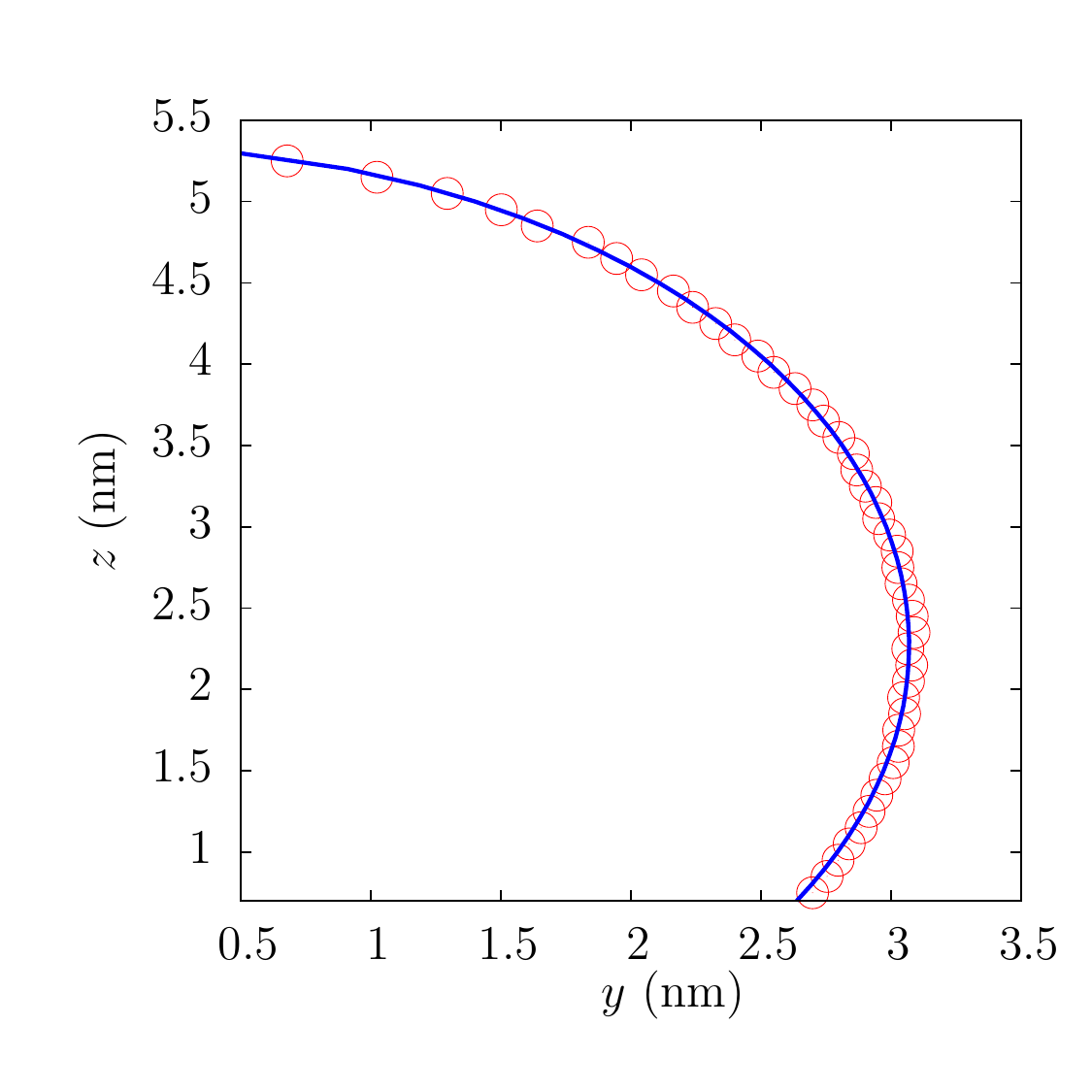}
\end{center}
\caption[Free Energy Derivatives]
{
(top) Snapshot of the cylindrical droplet used to calculate the contact angle.
(bottom) Fit of the average droplet profile (points) to Equation~\ref{eq:dropfit} (line).
}
\label{fig:ca}
\end{figure}

The contact angle was then determined by fitting this droplet profile to a circle for $z>0.7$~nm, as shown in Figure~\ref{fig:ca}, where the fit function is given by
\begin{equation}
y = \sqrt{b^2-\para{{z-a}}^2}.
\label{eq:dropfit}
\end{equation}
Using this functional form, the contact angle, $\theta$, can be obtained from
\begin{equation}
\frac{d y}{dz}\Bigg |_{z=0.7~{\rm nm}}  = -\frac{(z-a)}{\brac{b^2-(z-a)^2}^{1/2}} = \frac{1}{\tan(\pi-\theta)}.
\end{equation}
The contact angle was determined independently for each of the five slabs and averaged to yield a contact angle of $\theta=120.2\degree\pm 0.5\degree$,
or $\cos\theta=-0.503\pm0.008$.

%****************************************** New Section *********************************************************%
\section{Details of Vapor Basin Fits}

\subsection{Fitting used in the main text}

\begin{figure}[tb]
\begin{center}
\includegraphics[width=0.49\textwidth]{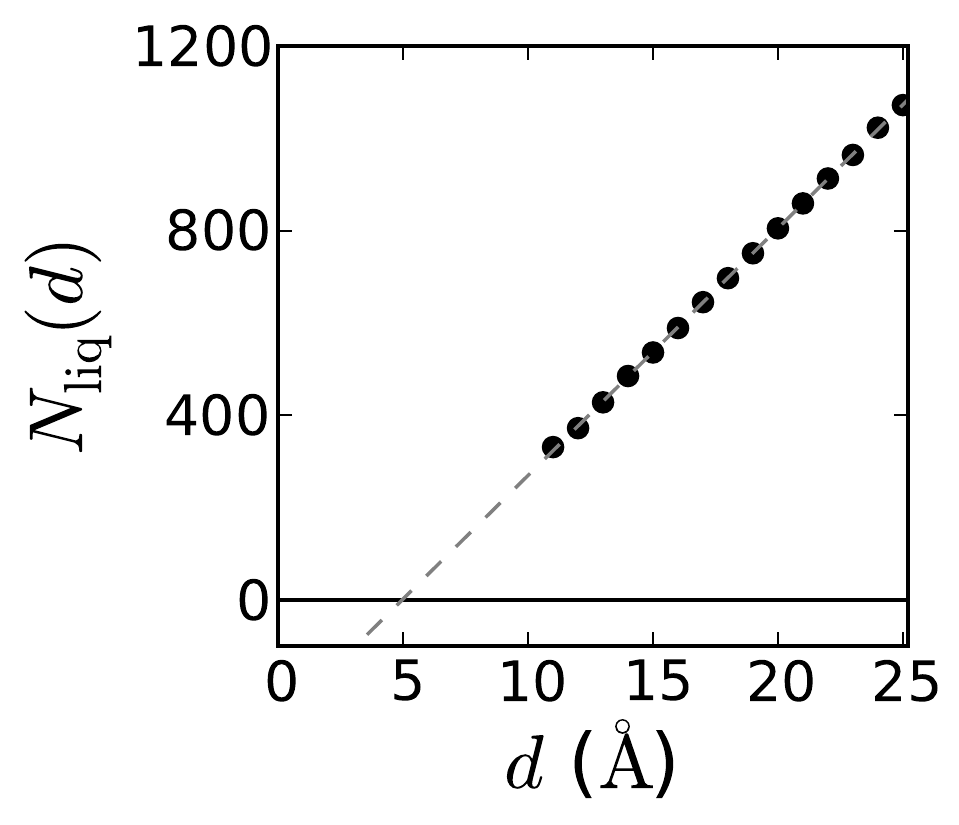}
\end{center}
\caption[Offset Determination]
{
The number of water molecules between the plates in the liquid basin, $N_{\rm liq}$, varies linearly with the distance between the plates, $d$, as measured between the centers of the atoms of the two plates. 
The effective distance between the plates, $d_{\rm eff}$, is obtained subtracting from $d$ the $x$-intercept, $\xi$, of a linear fit of $N_{\rm liq}(d)$; thus, $d_{\rm eff}=d-\xi$.
}
\label{fig:nliqD}
\end{figure}

In order to fit the simulated free energies to macroscopic theory, we consider
the formation of a vapor tube of radius $r$, which depends on the number of waters, $N$, between the hydrophobic plates as
%This will be true for $N<N_{\rm kink}$.
%Therefore, we consider writing 
%
\begin{equation}
r\approx \sqrt{\para{\frac{N_{\rm liq}-N}{N_{\rm liq}}} \frac{L^2}{\pi}}.
\label{eq:radN}
\end{equation}
%is the radius of the cylinder in terms of the number of waters between the plates.
Note that Equation~\ref{eq:radN} is an approximate expression;
%as there is not a one-to-one mapping between the number of waters between the plates and the radius of the vapor tube.
however, this simple approximation captures the salient features of vapor tube formation, as detailed below and in the main text.
The free energy as a function of $r$ as predicted by macroscopic theory is
\begin{equation}
\beta\Delta G(r) = 2\pi\beta\gamma \brac{r^2\cos\theta + r\para{d_{\rm eff} + \frac{2\lambda}{\gamma}}} 
- 2\ln(1-\frac{2r}{L}),
\label{eq:vapfit}
\end{equation}
where $\gamma$ is the liquid-vapor surface tension, $\theta$ is the contact angle (determined as described above), and $\lambda$ is the line tension.
The first term in Equation~\ref{eq:vapfit} is the free energy of vapor tube formation, $\Delta G_{\rm th}$, described in the main text, and the last term accounts for the translational entropy of the vapor tube.
%; note that the size of the plates appears only in this entropic term.
%
The effective distance between the plates, $d_{\rm eff}$, is obtained by subtracting a constant offset, $\xi$, from $d$, that is $d_{\rm eff}=d-\xi$.
The $x$-intercept of a linear fit of the simulated $d$-dependence of $N_{\rm liq}$ is equal to $\xi$, as shown in Figure~\ref{fig:nliqD}, so that a plot of $N_{\rm liq}$ vs $d_{\rm eff}$ passes through the origin.
%The effective distance between the plates is then given by $d_{\rm eff}=d-\xi$.
Here we find a value of the offset to be $\xi=0.4964$~nm.
The fits shown in Figure~4a of the main text were obtained by fitting the simulated free energies for $N<N_{\rm kink}$ and $r<L/2$ to the parameters, $\beta\gamma$ and $\lambda/\gamma$, using Equation~\ref{eq:vapfit}; the fit parameters obtained are listed in Table~\ref{tab:vapfit}.

%@@@@@@@@@@@@@@@@@@@@@@@@@@@@@@@@@@@
\begin{table}[t]
\caption{
\label{tab:vapfit}
Parameters obtained from fitting the vapor tube portion of the free energy to Equation~\ref{eq:vapfit}
}
\vspace{0.6cm}
\centering
\begin{tabular*}{\hsize}{@{\extracolsep{\fill}}lcc}
\hline
$d$ (\AA) & $\beta\gamma$ (nm$^{-2}$) & $\lambda/\gamma$ (nm) \\ 
\hline
11 & 15.1 & -0.150 \\[1.0ex] % $\pm$4\\

12 & 15.4 & -0.167 \\[1.0ex]

13 & 15.4 & -0.177 \\[1.0ex]

14 & 15.4 & -0.180 \\[1.0ex]

15 & 15.7 & -0.180 \\[1.0ex]

16 & 15.0 & -0.207 \\[1.0ex]

17 & 15.1 & -0.208 \\[1.0ex]

18 & 14.6 & -0.223 \\[1.0ex]

19 & 14.2 & -0.230 \\[1.0ex]

20 & 13.7 & -0.230 \\[1.0ex]

21 & 13.1 & -0.243 \\[1.0ex]

22 & 13.3 & -0.240 \\[1.0ex]

23 & 12.9 & -0.234 \\[1.0ex]

24 & 12.5 & -0.221 \\[1.0ex]

25 & 11.9 & -0.216 \\[1.0ex]

\end{tabular*}
\end{table}
%@@@@@@@@@@@@@@@@@@@@@@@@@@@@@@@@@@@

\subsection{Curvature Corrections}

We also explored a number of other possible fits, the first of which includes curvature corrections to the surface tension using the Tolman length, $\delta$,
such that the curvature corrected surface tension is $\gamma(r) = \gamma (1-\delta/r)$.
The fit equation then becomes
\begin{align}
\beta\Delta G(r) &= 2\pi\beta\gamma\brac{r^2\cos\theta + r\para{d_{\rm eff} + \frac{2\lambda}{\gamma} - \delta\cos\theta} - \delta d_{\rm eff}} \nonumber \\
&-2\ln\para{1-\frac{2r}{L}}.
\label{eq:tolman}
\end{align}
However, by referencing the simulated free energies for a given $d$-value to the free energy in the liquid basin, that is $\Delta G(N_{\rm liq})=0$, we force $\Delta G(r)$ to be 0 at $r=0$.
To be consistent with this convention, we neglect the constant term, $2\pi\gamma\delta d_{\rm eff}$ in Equation~\ref{eq:tolman}, and instead fit to
\begin{align}
\beta\Delta G(r) &= 2\pi\beta\gamma\brac{r^2\cos\theta + r\para{d_{\rm eff} + \frac{2\lambda}{\gamma} - \delta\cos\theta}} \nonumber \\
&-2\ln\para{1-\frac{2r}{L}}.
\label{eq:tolman2}
\end{align}
%We note that determining the free energy differences between $d$-values would enable an estimation of $\delta$, and therefore the constant term in Equation~\ref{eq:tolman}. However, this is beyond the scope of the current work.

From Equation~\ref{eq:tolman2}, we see that the curvature correction $\delta$ acts in a manner analogous to the line tension $\lambda$,
\ie \ both are coefficients in the term linear in $r$. 
Therefore, we can simply relate the fit parameters in Equation~\ref{eq:tolman2} to those in Equation~\ref{eq:vapfit}, according to
\begin{equation}
\frac{\lambda}{\gamma} = \frac{\lambda'}{\gamma'}+\frac{\delta}{2}\cos\theta,
\end{equation}
where the primed quantities indicate those obtained from Equation~\ref{eq:vapfit}.
The value of $\lambda/\gamma$ obtained from Equation~\ref{eq:tolman2} can then be readily predicted with knowledge of the Tolman length.
Recent estimates of this length for SPC/E water at 300~K yield $\delta\approx-0.1$~nm~\cite{Sedlmeier:2012aa,Suri:PRL:2014}. 
Therefore, the value of $\lambda/\gamma$ changes by ten to sixteen percent through the inclusion of the Tolman length in our fitting procedure.

\subsection{Omission of Line Tension}

To ascertain the importance of line tension, we attempt to fit the simulated free energies without including the term containing $\lambda$ in our expression for the macroscopic theory, and instead using $\delta$ as a fit parameter
according to
\begin{align}
\beta\Delta G(r) &= 2\pi\beta\gamma\brac{r^2\cos\theta + r\para{d_{\rm eff} - \delta\cos\theta} } \nonumber \\
&-2\ln\para{1-\frac{2r}{L}}.
\label{eq:noline}
\end{align}
As one may anticipate from the above discussion, the data is fit equally well by Equation~\ref{eq:noline}.
However, this procedure yields unphysical estimates of the Tolman length, with $\delta\sim -1$~nm.
Therefore, we can conclude that line tension is necessary to provide a physically accurate description
of the vapor tube formation and growth that facilitates capillary evaporation between nanoscopic hydrophobic plates.
%, in accord with previous findings~\cite{Sharma:2012aa}.

%****************************************** New Section *********************************************************%
\section{Liquid Basin Fits}
\label{sec:liqfits}

\begin{figure}[tb]
\begin{center}
\includegraphics[width=0.45\textwidth]{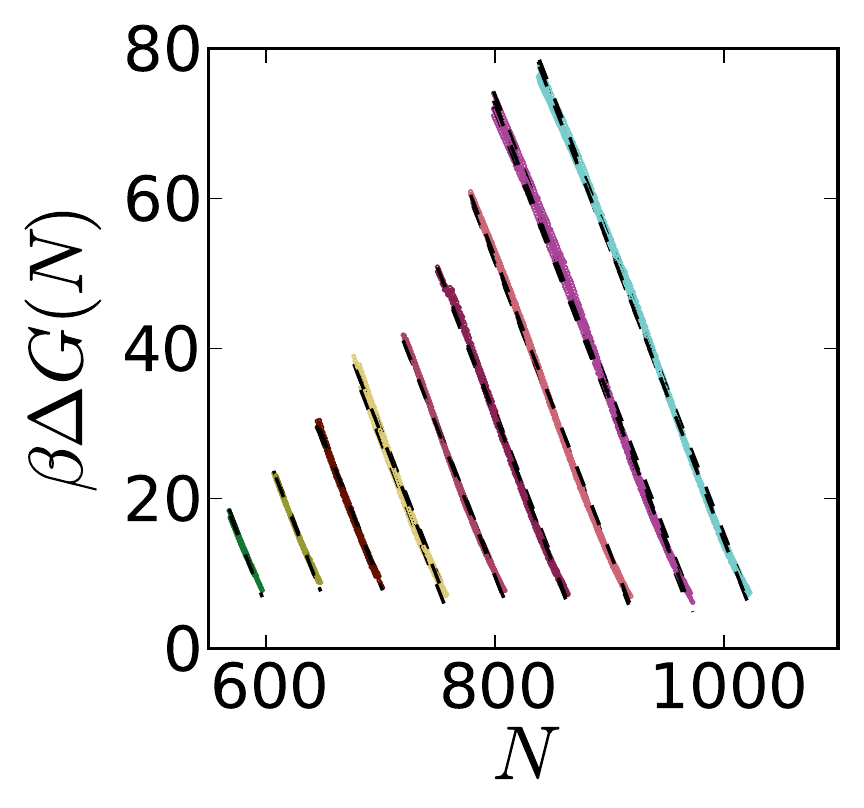}
\includegraphics[width=0.45\textwidth]{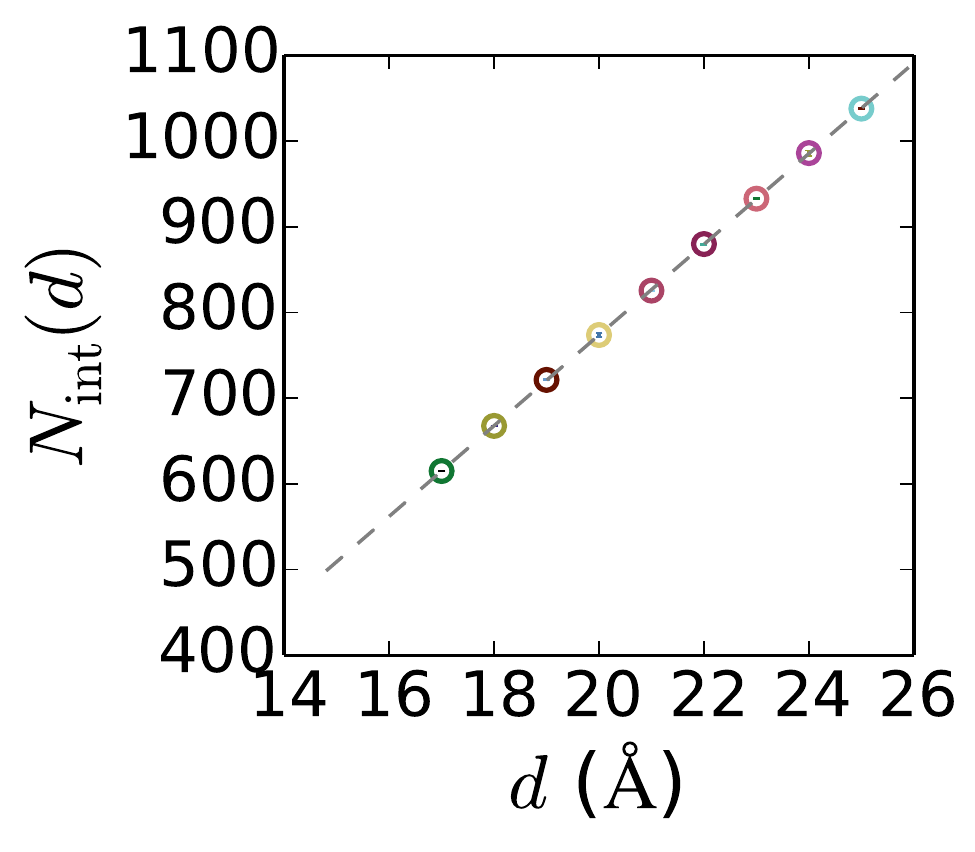}
\end{center}
\caption[Liquid Fitting]
{
(top) Portions of the free energies that were linearly fit for $d>16$~\AA. 
Simulation data, split into five blocks for error estimation, is shown as colored data points.
Linear fits are shown as black dashed lines.
(bottom) The $x$-intercept, $N_{\rm int}$, obtained from the linear fits, is plotted as a function of $d$.
}
\label{fig:liqfit}
\end{figure}

%\subsection{Formation of Bubbles}
In order to characterize the free energetics of isolated cavities, 
 %corresponding to the fat tails at low $N$ in the liquid basin, 
we fit the simulated free energies in the range, $N_{\rm kink}+20 < N < N_{\rm liq}-50$, to straight lines.
We first fit the $d=25$~\AA \ free energy, and use the corresponding slope for all $d$-values.
Intercepts are then obtained by fitting the data for each $d$-value separately.
For $d\le16$~\AA, data in the $N_{\rm kink}+20 < N < N_{\rm liq}-50$-range are insufficient to be fit reliably, so the fitting was not performed.
The regions fit for each $d$-value are shown in Figure~\ref{fig:liqfit}. 
In order to estimate errors, we divided the data into five blocks of equal length, and fit each block separately (similar block averaging analysis was performed for all quantities). 
Therefore, for each $d$, we show five data sets and five fits, although they are nearly indistinguishable.

\begin{figure}[tb]
\begin{center}
\includegraphics[width=0.49\textwidth]{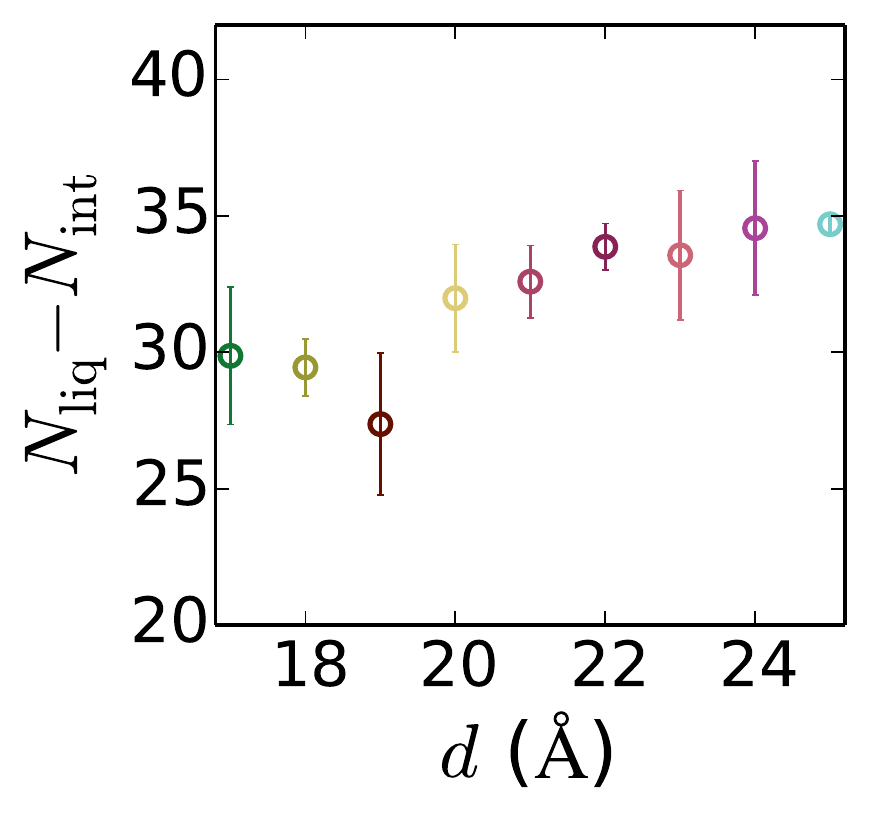}
\end{center}
\caption[Liquid Fit Analysis]
{
The number of waters needed to be removed from the liquid basin to encroach on the region of the fat tail,
as measured by $N_{\rm liq}-N_{\rm int}$ for $d>16$~\AA.
Error bars correspond to one standard deviation.
}
\label{fig:nliqNint}
\end{figure}

From these fits, we can examine the behavior of $N_{\rm liq}(d)-N_{\rm int}(d)$, where $N_{\rm int}(d)$ is the $x$-intercept of the linear fit to the fat tails (for $d>16$~\AA).
Both $N_{\rm liq}(d)$ and $N_{\rm int}(d)$ are roughly linear in $d$, see Figures~\ref{fig:nliqD} and~\ref{fig:liqfit}, respectively.
However, as shown in Figure~\ref{fig:nliqNint}, the quantity $N_{\rm liq}-N_{\rm int}$, which represents the distance in $N$ from the liquid basin one must move to observe non-Gaussian tails in $\Delta G(N;d)$, or equivalently, to form isolated cavities, is largely insensitive to $d$.

%\subsection{Estimating Bubble Height}
%The linear dependence of the free energy in the fat tail region is consistent with the growth of a cylindrical, disc-like bubble growing with fixed height, $H$. 
%Through consideration of the contact between this vapor bubble and the surface, we can approximate its height by
%$H\approx \beta\gamma\cos\theta/m\rho^{\rm bulk}_{\rm water}$, where $m=\partial \beta\Delta G/\partial N$ is the slope obtained from linear fitting of the fat low-$N$ tail.
%This procedure yields a height of $H\approx0.5$~nm for a vapor bubble adjacent to the plate for all $d$-values. 

%****************************************** New Section *********************************************************%

\section{Barrier Location and Height: Simulations vs Macroscopic Theory}

\begin{figure}[tb]
\begin{center}
\includegraphics[width=0.4\textwidth]{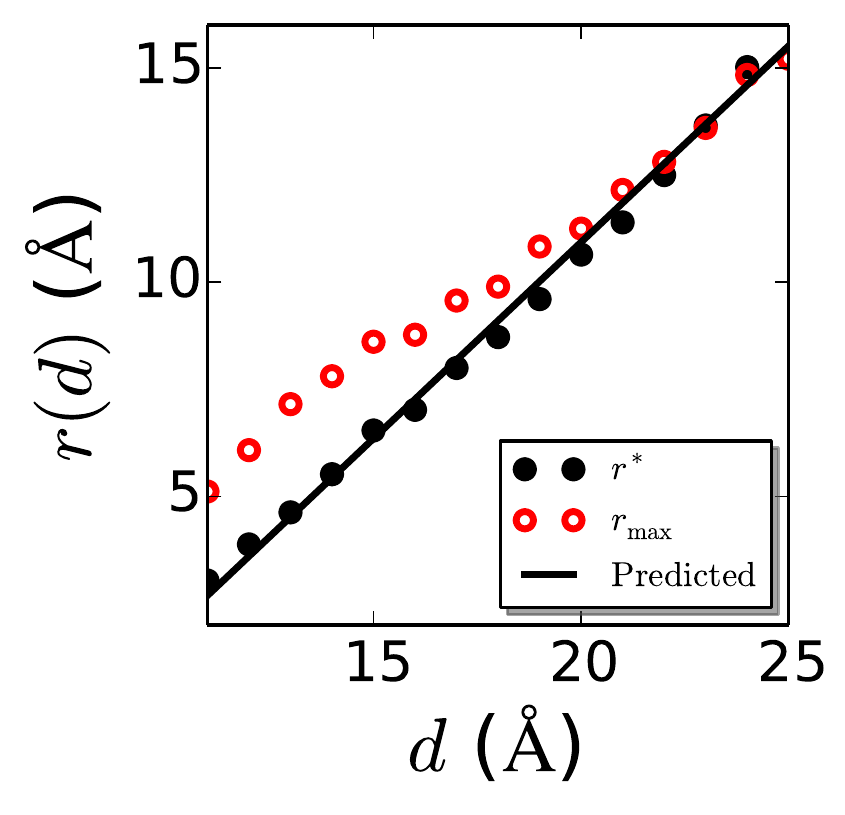}
\includegraphics[width=0.45\textwidth]{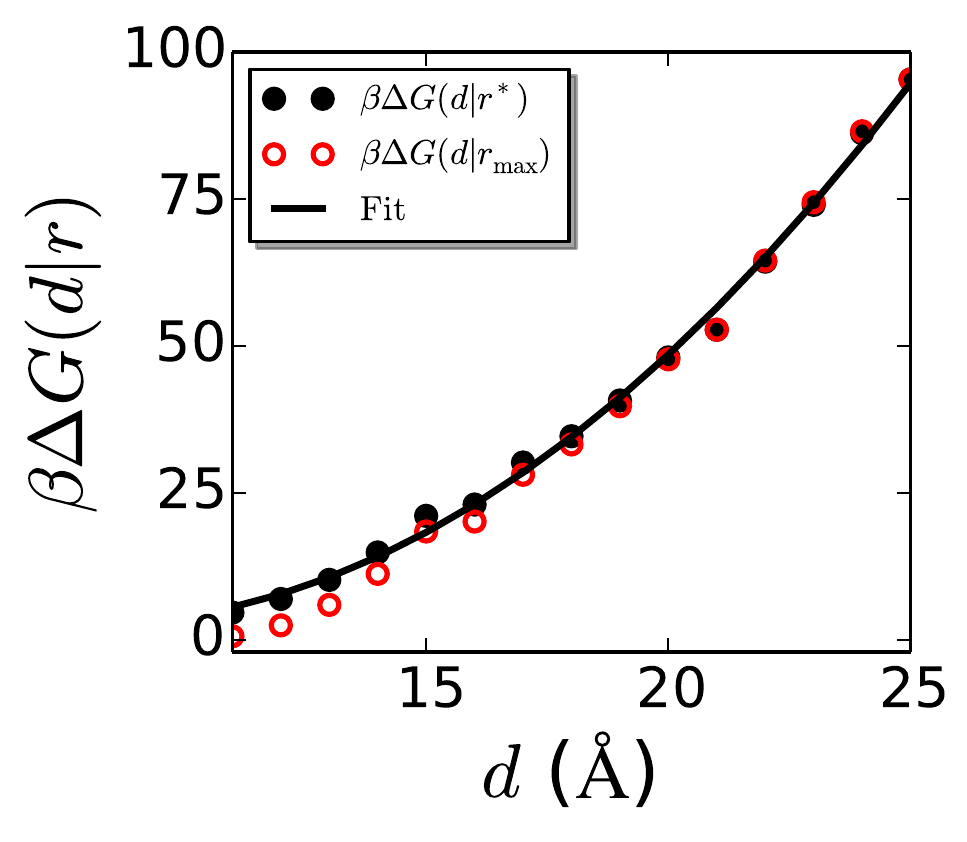}
\end{center}
\caption[Barrier Analysis]
{
(top) Barrier position $r^*$ and (bottom) barrier height $\beta\Delta G(r^*)$ of the fitted vapor tube free energies, are well-fit by linear and parabolic functions of the inter-plate separation, $d$, respectively, in agreement with macroscopic theory.
In contrast, the simulated barrier position $r_{\rm max}$ and barrier height $\beta\Delta G(r_{\rm max})$ deviate from those functional forms.
}
\label{fig:rstar}
\end{figure}

In this section, we focus on the location as well as the height of the barrier, as predicted by macroscopic theory, and compare them with the corresponding simulated values.
%as described by Equation~\ref{eq:vapfit}. 
%The location, $r^*(d)$, and height, $\Delta G(d|r^*)$ of these barriers as a function of $d$ are shown in Figure~\ref{fig:rstar}, as well as a simple linear fit to the simulation data.
%as well as the $d$-dependence predicted by macroscopic theory using the average of $\gamma$ and $\lambda$ over all $d$-values
%and a simple linear fit to the simulation data.
Macroscopic theory predicts a linear dependence of the location of the barrier, $r^*$, on $d$,
\begin{equation}
r^*(d) = - \frac{d_{\rm eff}}{2\cos\theta} - \frac{\lambda}{\gamma \cos\theta},
\end{equation}
where we have neglected the logarithmic term in Equation~\ref{eq:vapfit}.
%\begin{equation}
%r^*(d) = \frac{AL-B+\sqrt{(AL)^2 + 2ABL + 16A+B^2} }{4A},
%\label{eq:rstar}
%\end{equation}
%where $A\equiv 2\pi\gamma\cos\theta$ and $B\equiv 2\pi\gamma(d_{\rm eff} + 2\lambda/\gamma)$.
This behavior is captured by the data shown in Figure~\ref{fig:rstar}, where the $r^*$-values are obtained from the fits of the vapor tubes free energetics to macroscopic theory.
%Equation~\ref{eq:rstar} slightly overestimates the magnitudes of the slope and intercept, as compared with the linear regression, shown as a solid line.
The slope of the fitted line corresponds to $\cos\theta=-0.547$, which is slightly larger in magnitude than that obtained from direct simulations.
The ratio $\lambda/\gamma=-0.16$~nm obtained from the $y$-intercept of the linear fit lies in the same range as those predicted from fitting the free energies to Equation~\ref{eq:vapfit}.
Macroscopic theory also predicts $\Delta G(r^*;d)$ to be quadratic in $d$; 
%(with the added natural logarithm term due to entropic effects).
this behavior is similarly captured %but again the parameters are overestimated
as illustrated by fitting the data to a parabola (solid line).
%XXX expression; parameters extracted; how do they compare to previously obtained fit parameters?

In addition to the position and height of the maximum of the fits, we also include the position and the height of the free energy barrier obtained from the simulated free energy profiles, $r_{\rm max}(d)$ and $\Delta G(r_{\rm max};d)$, respectively.
For large $d$-values, we expect $r_{\rm max}\approx r^*$, however, this is not true at smaller plate separations because the critical vapor tube is in the metastable branch of the vapor tube free energy; for these $d$-values, the barrier corresponds to the kink in the free energy.
Indeed, $r^*\neq r_{\rm max}$ at these separations, and the dependence of $r_{\rm max}$ on $d$ can not be described by a straight line over the entire range of $d$.
Similar behavior is observed in $\Delta G(r_{\rm max}; d)$, albeit to a lesser extent.

%****************************************** New Section *********************************************************%
\section{Data for All Systems Studied}

\begin{figure}[tb]
\begin{center}
\includegraphics[width=0.45\textwidth]{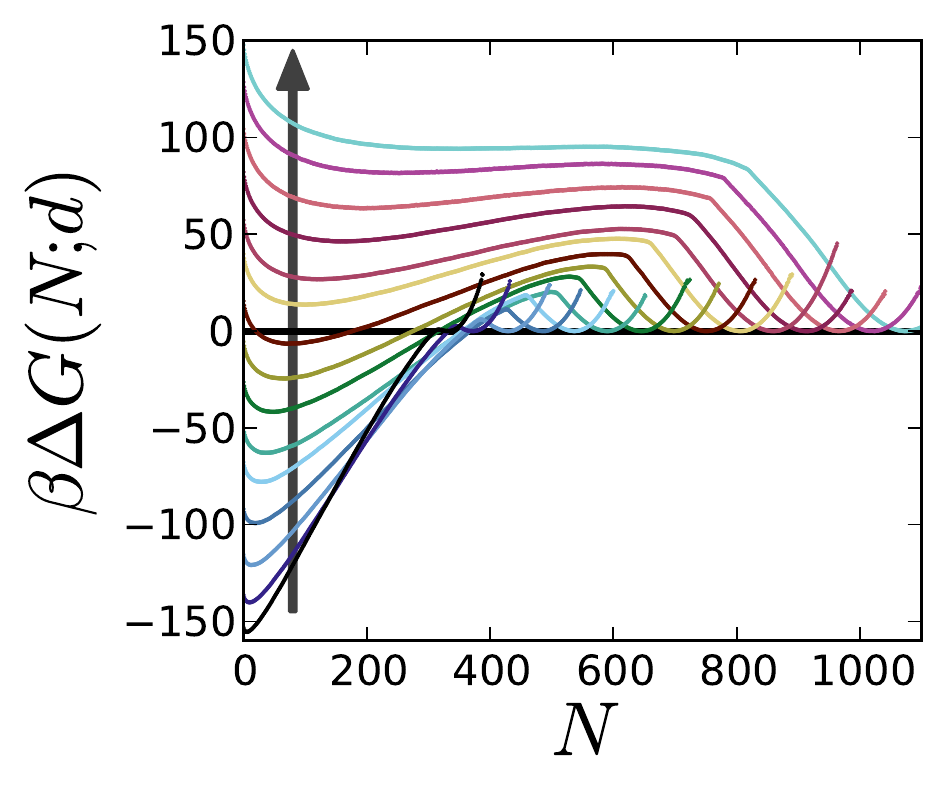}
\includegraphics[width=0.45\textwidth]{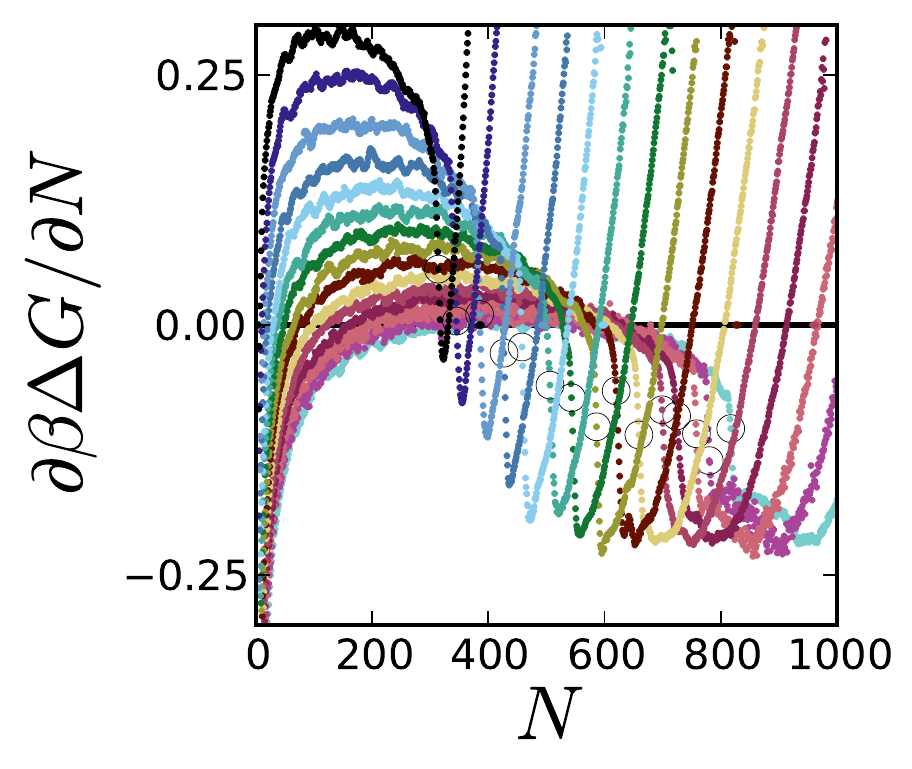}
\end{center}
\caption[Free Energy Derivatives]
{
The dependence of (top) the free energies and (bottom) their smoothed derivatives on $N$ for all $d$-values studied.
Arrows point in the direction of increasing $d$, from $d=11$~\AA \ to $d=25$~\AA \ in increments of 1~\AA.
}
\label{fig:dfeall}
\end{figure}

\begin{figure}[tb]
\begin{center}
\includegraphics[width=0.45\textwidth]{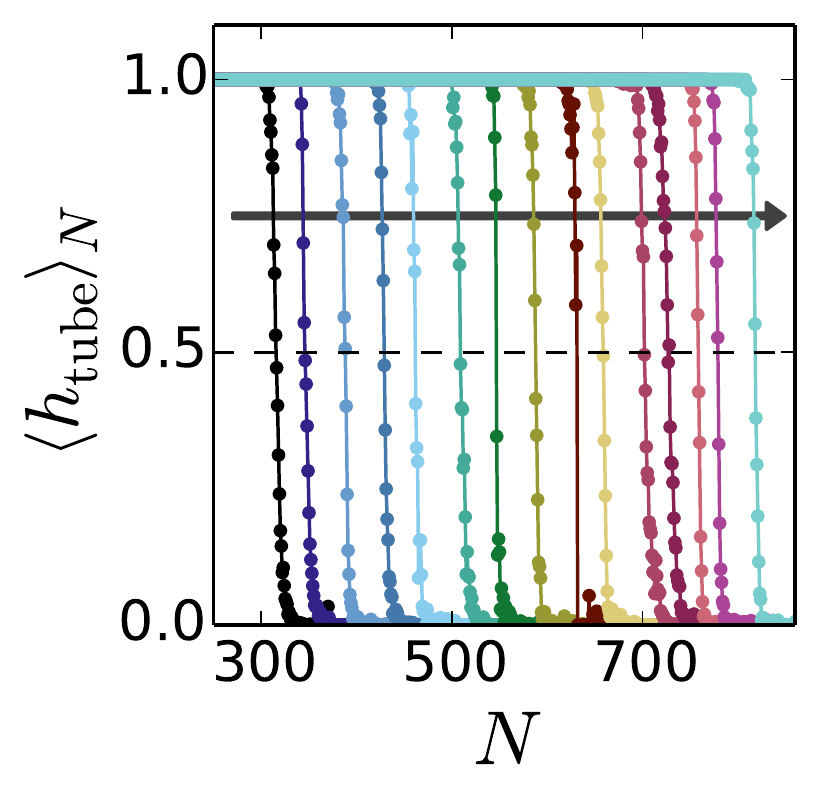}
\end{center}
\caption[Tube Indicator Function]
{
$\avg{h_{\rm tube}}_N$ for all $d$-values studied.
Arrows point in the direction of increasing $d$, from $d=11$~\AA \ to $d=25$~\AA \ in increments of 1~\AA.
}
\label{fig:allhtube}
\end{figure}

\begin{figure}[tb]
\begin{center}
\includegraphics[width=0.45\textwidth]{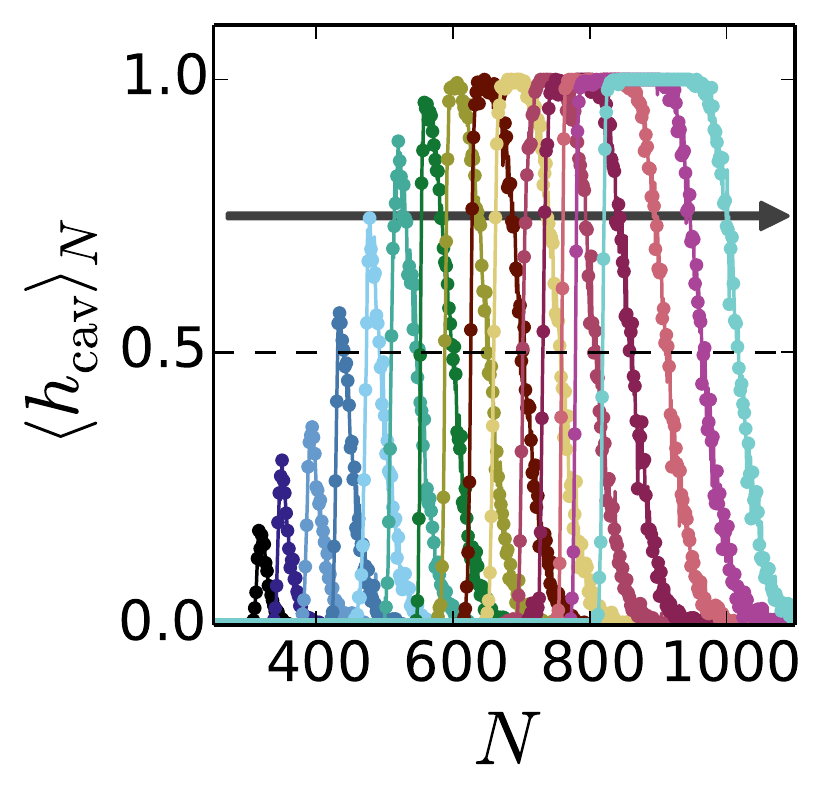}
\end{center}
\caption[Cavity Indicator Function]
{
$\avg{h_{\rm cav}}_N$ for all $d$-values studied.
Arrows point in the direction of increasing $d$, from $d=11$~\AA \ to $d=25$~\AA \ in increments of 1~\AA.
}
\label{fig:allhcav}
\end{figure}

\begin{figure}[tb]
\begin{center}
\includegraphics[width=0.45\textwidth]{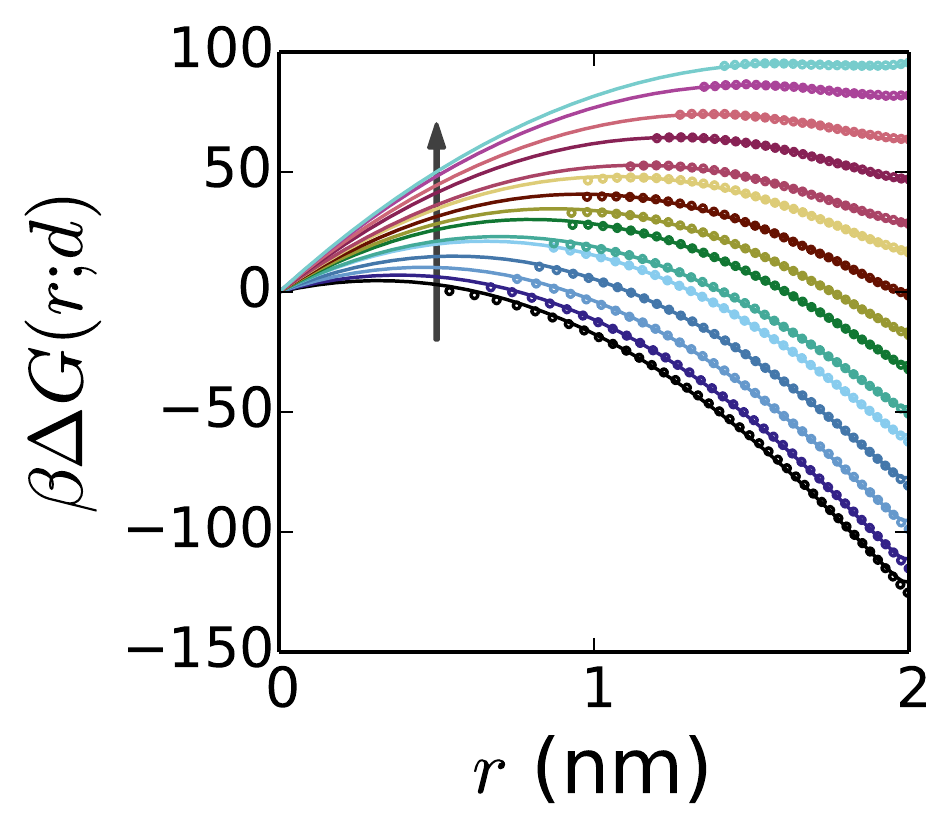}
\end{center}
\caption[Vapor Basin Fits]
{
Free energies in the vapor basin ($N<N_{\rm kink}$) and the corresponding fits to Equation~\ref{eq:vapfit}.
Simulation data is shown as points and the fits are shown as solid lines. 
The arrow indicates the direction of increasing $d$, from $d=11$~\AA \ to $d=25$~\AA \ in increments of 1~\AA.
}
\label{fig:vapalld}
\end{figure}

\begin{figure}[tb]
\begin{center}
\includegraphics[width=0.45\textwidth]{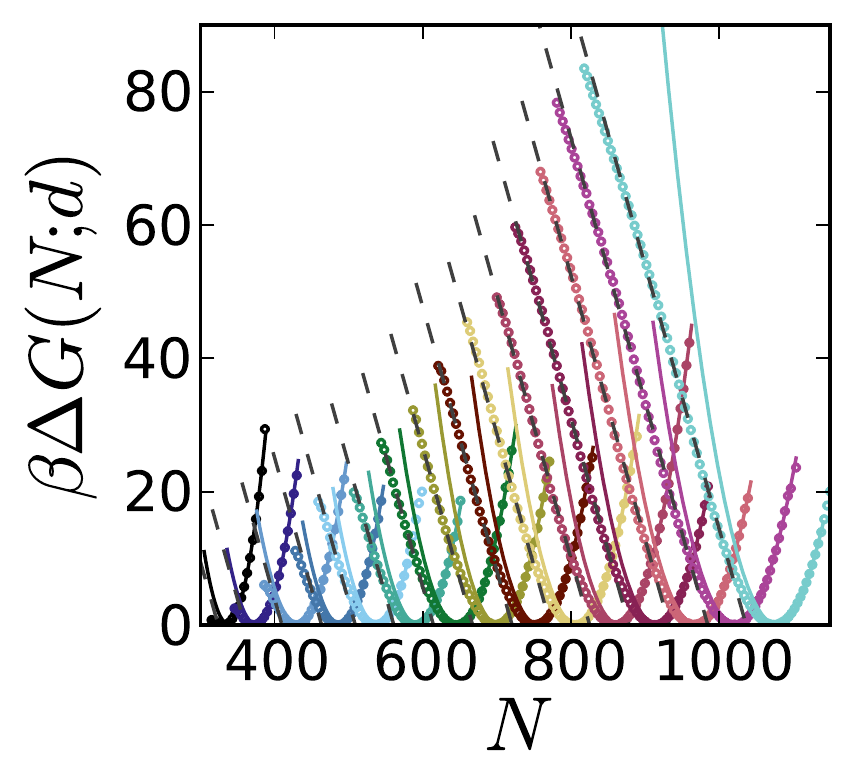}
\end{center}
\caption[Liquid Basin Fits]
{
Free energies in the liquid basin ($N>N_{\rm kink}$) and the corresponding parabolic (solid) and linear (dashed) fits 
to the right side of the minimum and to the fat tail regions, respectively.
}
\label{fig:liqalld}
\end{figure}

For completeness, we include the free energies, their derivatives, $\avg{h_{\rm tube}}_N$, and $\avg{h_{\rm cav}}_N$, as well as fits to the vapor and liquid basins, for all $d$-values studied in Figures~\ref{fig:dfeall},~\ref{fig:allhtube},~\ref{fig:allhcav},~\ref{fig:vapalld}, and~\ref{fig:liqalld} respectively.
%Additionally, we include the vapor basin fits and linear regions (where applicable) for all $d$-values in Figure~\ref{fig:alld}.

%\end{article}

%@@@@@@@@@@@@@@@@@@@@@@@@@@@@@@@@@@@@@@@@@@@@@@
%@@@@@@@@@@@@@@@@@@@@@@@@@@@@@@@@@@@@@@@@@@@@@@

%@@@@@@@@@@@@@@@@@@@@@@@@@@@@@@@@@@@@@@@@@@@@@@
%@@@@@@@@@@@@@@@@@@@@@@@@@@@@@@@@@@@@@@@@@@@@@@

%@@@@@@@@@@@@@@@@@@@@@@@@@@@@@@@@@@@@@@@@@@@@@@
%@@@@@@@@@@@@@@@@@@@@@@@@@@@@@@@@@@@@@@@@@@@@@@

%@@@@@@@@@@@@@@@@@@@@@@@@@@@@@@@@@@@@@@@@@@@@@@
%@@@@@@@@@@@@@@@@@@@@@@@@@@@@@@@@@@@@@@@@@@@@@@

%@@@@@@@@@@@@@@@@@@@@@@@@@@@@@@@@@@@@@@@@@@@@@@
%@@@@@@@@@@@@@@@@@@@@@@@@@@@@@@@@@@@@@@@@@@@@@@

%@@@@@@@@@@@@@@@@@@@@@@@@@@@@@@@@@@@@@@@@@@@@@@
%@@@@@@@@@@@@@@@@@@@@@@@@@@@@@@@@@@@@@@@@@@@@@@

%@@@@@@@@@@@@@@@@@@@@@@@@@@@@@@@@@@@@@@@@@@@@@@
%@@@@@@@@@@@@@@@@@@@@@@@@@@@@@@@@@@@@@@@@@@@@@@

%@@@@@@@@@@@@@@@@@@@@@@@@@@@@@@@@@@@@@@@@@@@@@@
%@@@@@@@@@@@@@@@@@@@@@@@@@@@@@@@@@@@@@@@@@@@@@@

%@@@@@@@@@@@@@@@@@@@@@@@@@@@@@@@@@@@@@@@@@@@@@@
%@@@@@@@@@@@@@@@@@@@@@@@@@@@@@@@@@@@@@@@@@@@@@@

%@@@@@@@@@@@@@@@@@@@@@@@@@@@@@@@@@@@@@@@@@@@@@@
%@@@@@@@@@@@@@@@@@@@@@@@@@@@@@@@@@@@@@@@@@@@@@@

%@@@@@@@@@@@@@@@@@@@@@@@@@@@@@@@@@@@@@@@@@@@@@@
%@@@@@@@@@@@@@@@@@@@@@@@@@@@@@@@@@@@@@@@@@@@@@@

%@@@@@@@@@@@@@@@@@@@@@@@@@@@@@@@@@@@@@@@@@@@@@@
%@@@@@@@@@@@@@@@@@@@@@@@@@@@@@@@@@@@@@@@@@@@@@@

%@@@@@@@@@@@@@@@@@@@@@@@@@@@@@@@@@@@@@@@@@@@@@@
%@@@@@@@@@@@@@@@@@@@@@@@@@@@@@@@@@@@@@@@@@@@@@@

%@@@@@@@@@@@@@@@@@@@@@@@@@@@@@@@@@@@@@@@@@@@@@@
%@@@@@@@@@@@@@@@@@@@@@@@@@@@@@@@@@@@@@@@@@@@@@@

%@@@@@@@@@@@@@@@@@@@@@@@@@@@@@@@@@@@@@@@@@@@@@@
%@@@@@@@@@@@@@@@@@@@@@@@@@@@@@@@@@@@@@@@@@@@@@@

%@@@@@@@@@@@@@@@@@@@@@@@@@@@@@@@@@@@@@@@@@@@@@@
%\begin{figure*}[tb]
%\begin{center}
%\includegraphics[width=0.25\textwidth]{d11Zoom-ForPaper.pdf}
%\includegraphics[width=0.25\textwidth]{d12Zoom-ForPaper.pdf}
%\includegraphics[width=0.25\textwidth]{d13Zoom-ForPaper.pdf}
%\includegraphics[width=0.25\textwidth]{d14Zoom-ForPaper.pdf}
%\includegraphics[width=0.25\textwidth]{d15Zoom-ForPaper.pdf}
%\includegraphics[width=0.25\textwidth]{d16Zoom-ForPaper.pdf}
%\includegraphics[width=0.25\textwidth]{d17Zoom-ForPaper.pdf}
%\includegraphics[width=0.25\textwidth]{d18Zoom-ForPaper.pdf}
%\includegraphics[width=0.25\textwidth]{d19Zoom-ForPaper.pdf}
%\includegraphics[width=0.25\textwidth]{d20Zoom-ForPaper.pdf}
%\includegraphics[width=0.25\textwidth]{d21Zoom-ForPaper.pdf}
%\includegraphics[width=0.25\textwidth]{d22Zoom-ForPaper.pdf}
%\includegraphics[width=0.25\textwidth]{d23Zoom-ForPaper.pdf}
%\includegraphics[width=0.25\textwidth]{d24Zoom-ForPaper.pdf}
%\includegraphics[width=0.25\textwidth]{d25Zoom-ForPaper.pdf}
%\end{center}
%\caption[Free Energy Derivatives]
%{
%Vapor basin fits to Equation~\ref{eq:vapfit} (dot-dashed) and liquid basin linear fits (dashed) near $N_{\rm kink}$ for all $d$-values studied.
%}
%\label{fig:alld}
%\end{figure*}
%@@@@@@@@@@@@@@@@@@@@@@@@@@@@@@@@@@@@@@@@@@@@@@

%% file: MakeArxivMerged.bbl
\begin{thebibliography}{10}

\bibitem{tanford_book}
Tanford C
\newblock (1973) \emph{The Hydrophobic Effect - Formation of Micelles and
  Biological Membranes}
\newblock (Wiley Interscience, New York).

\bibitem{kauzmann}
Kauzmann W
\newblock (1959) Some factors in the interpretation of protein denaturation.
\newblock \emph{Adv. Prot. Chem.} 14:1--63.

\bibitem{FHS:1973}
Stillinger FH
\newblock (1973) Structure in aqueous solutions of nonpolar solutes from the
  standpoint of scaled-particle theory.
\newblock \emph{J. Solution Chem.} 2:141--158.

\bibitem{Israelachvili_book}
Israelachvili JN
\newblock (2011) \emph{Intermolecular and surface forces: revised third
  edition}
\newblock (Academic press).

\bibitem{Chandler:Nature:2005}
Chandler D
\newblock (2005) Interfaces and the driving force of hydrophobic assembly.
\newblock \emph{Nature} 437:640--647.

\bibitem{Israelachvili:Pashley:1982}
Israelachvili J, Pashley R
\newblock (1982) The hydrophobic interaction is long range, decaying
  exponentially with distance.
\newblock \emph{Nature} 300:341 -- 342.

\bibitem{Christenson:Science:1988}
Christenson HK, Claesson PM
\newblock (1988) Cavitation and the interaction between macroscopic hydrophobic
  surfaces.
\newblock \emph{Science} 239:390--392.

\bibitem{Berard:JCP:1993}
B{\'e}rard DR, Attard P, Patey GN
\newblock (1993) {Cavitation of a Lennard-Jones fluid between hard walls, and
  the possible relevance to the attraction measured between hydrophobic
  surfaces}.
\newblock \emph{J. Chem. Phys.} 98:7236--7244.

\bibitem{Parker:JPC:1994}
Parker JL, Claesson PM, Attard P
\newblock (1994) Bubbles, cavities, and the long-ranged attraction between
  hydrophobic surfaces.
\newblock \emph{J. Phys. Chem.} 98:8468--8480.

\bibitem{wallqvist2}
Wallqvist A, Berne BJ
\newblock (1995) Computer simulation of hydrophobic hydration forces on stacked
  plates at short range.
\newblock \emph{J. Phys. Chem.} 99:2893--2899.

\bibitem{Lum:PRE:1997}
Lum K, Luzar A
\newblock (1997) Pathway to surface-induced phase transition of a confined
  fluid.
\newblock \emph{Phys Rev E} 56:R6283.

\bibitem{Lum:IntJT:1998}
Lum K, Chandler D
\newblock (1998) Phase diagram and free energies of vapor films and tubes for a
  confined fluid.
\newblock \emph{Intl J Thermophys} 19:845--855.

\bibitem{LCW}
Lum K, Chandler D, Weeks JD
\newblock (1999) Hydrophobicity at small and large length scales.
\newblock \emph{J. Phys. Chem. B} 103:4570--4577.

\bibitem{Bolhuis:JCP:2000}
Bolhuis PG, Chandler D
\newblock (2000) Transition path sampling of cavitation between molecular scale
  solvophobic surfaces.
\newblock \emph{J. Chem. Phys.} 113:8154--8160.

\bibitem{Leung:PRL:2003}
Leung K, Luzar A, Bratko D
\newblock (2003) Dynamics of capillary drying in water.
\newblock \emph{Phys. Rev. Lett.} 90:065502.

\bibitem{HuangX:PNAS:2003}
Huang X, Margulis CJ, Berne BJ
\newblock (2003) Dewetting-induced collapse of hydrophobic particles.
\newblock \emph{Proc Natl Acad Sci USA} 100:11953--11958.

\bibitem{urbic2006confined}
Urbic T, Vlachy V, Dill KA
\newblock (2006) Confined water: a mercedes-benz model study.
\newblock \emph{J Phys Chem B} 110:4963--4970.

\bibitem{Choudhury:JACS:2007}
Choudhury N, Pettitt BM
\newblock (2007) The dewetting transition and the hydrophobic effect.
\newblock \emph{J. Am. Chem. Soc.} 129:4847--4852.

\bibitem{Xu:JPCB:2010}
Xu L, Molinero V
\newblock (2010) Liquid- vapor oscillations of water nanoconfined between
  hydrophobic disks: Thermodynamics and kinetics.
\newblock \emph{J Phys Chem B} 114:7320--7328.

\bibitem{Sharma:PNAS:2012}
Sharma S, Debenedetti PG
\newblock (2012) Evaporation rate of water in hydrophobic confinement.
\newblock \emph{Proc Natl Acad Sci USA} 109:4365--4370.

\bibitem{Sharma:JPCB:2012}
Sharma S, Debenedetti PG
\newblock (2012) Free energy barriers to evaporation of water in hydrophobic
  confinement.
\newblock \emph{J Phys Chem B} 116:13282--13289.

\bibitem{Ducker:PRL:2012}
Mastropietro DJ, Ducker WA
\newblock (2012) Forces between hydrophobic solids in concentrated aqueous salt
  solution.
\newblock \emph{Phys. Rev. Lett.} 108:106101.

\bibitem{Hagan:2012:Capsid}
Yu N, Hagan MF
\newblock (2012) {Simulations of HIV capsid protein dimerization reveal the
  effect of chemistry and topography on the mechanism of hydrophobic protein
  association}.
\newblock \emph{Biophys J} 103:1363--1369.

\bibitem{Discher:Science:1999}
Discher BM, {et~al.}
\newblock (1999) Polymersomes: tough vesicles made from diblock copolymers.
\newblock \emph{Science} 284:1143--1146.

\bibitem{Percec:Science:2010}
Percec V, {et~al.}
\newblock (2010) Self-assembly of janus dendrimers into uniform dendrimersomes
  and other complex architectures.
\newblock \emph{Science} 328:1009--1014.

\bibitem{Janus:Granick}
Chen Q, Yan J, Zhang J, Bae SC, Granick S
\newblock (2012) Janus and multiblock colloidal particles.
\newblock \emph{Langmuir} 28:13555--13561.

\bibitem{Vargo:PNAS:2012}
Vargo KB, Parthasarathy R, Hammer DA
\newblock (2012) Self-assembly of tunable protein suprastructures from
  recombinant oleosin.
\newblock \emph{Proc Natl Acad Sci USA} 109:11657--11662.

\bibitem{Berne:ARPC:2009}
Berne BJ, Weeks JD, Zhou R
\newblock (2009) Dewetting and hydrophobic interaction in physical and
  biological systems.
\newblock \emph{Annu Rev Phys Chem} 60:85--103.

\bibitem{PGD:JPCL:2011}
Cerdeirina CA, Debenedetti PG, Rossky PJ, Giovambattista N
\newblock (2011) Evaporation length scales of confined water and some common
  organic liquids.
\newblock \emph{J. Phys. Chem. Lett.} 2:1000--1003.

\bibitem{Giovambattista:ARPC:2012}
Giovambattista N, Rossky P, Debenedetti P
\newblock (2012) Computational studies of pressure, temperature, and surface
  effects on the structure and thermodynamics of confined water.
\newblock \emph{Annu Rev Phys Chem} 63:179--200.

\bibitem{Patel:JPCB:2010}
Patel AJ, Varilly P, Chandler D
\newblock (2010) Fluctuations of water near extended hydrophobic and
  hydrophilic surfaces.
\newblock \emph{J. Phys. Chem. B} 114:1632 -- 1637.

\bibitem{Patel:JSP:2011}
Patel AJ, Varilly P, Chandler D, Garde S
\newblock (2011) Quantifying density fluctuations in volumes of all shapes and
  sizes using indirect umbrella sampling.
\newblock \emph{J. Stat. Phys.} 145:265 -- 275.

\bibitem{Hummer:PNAS:1996}
Hummer G, Garde S, Garcia AE, Pohorille A, Pratt LR
\newblock (1996) An information theory model of hydrophobic interactions.
\newblock \emph{Proc Natl Acad Sci USA} 93:8951--8955.

\bibitem{Garde:PRL:1996}
Garde S, Hummer G, Garcia AE, Paulaitis ME, Pratt LR
\newblock (1996) Origin of entropy convergence in hydrophobic hydration and
  protein folding.
\newblock \emph{Phys. Rev. Lett.} 77:4966--4968.

\bibitem{Rajamani:PNAS:2005}
Rajamani S, Truskett TM, Garde S
\newblock (2005) Hydrophobic hydration from small to large lengthscales:
  Understanding and manipulating the crossover.
\newblock \emph{Proc Natl Acad Sci USA} 102:9475--9480.

\bibitem{Mittal:PNAS:2008}
Mittal J, Hummer G
\newblock (2008) Static and dynamic correlations in water at hydrophobic
  interfaces.
\newblock \emph{Proc Natl Acad Sci USA} 105:20130--20135.

\bibitem{Godawat:PNAS:2009}
Godawat R, Jamadagni SN, Garde S
\newblock (2009) Characterizing hydrophobicity of interfaces by using cavity
  formation, solute binding, and water correlations.
\newblock \emph{Proc Natl Acad Sci USA} 106:15119 -- 15124.

\bibitem{LLCW}
Varilly P, Patel AJ, Chandler D
\newblock (2011) An improved coarse-grained model of solvation and the
  hydrophobic effect.
\newblock \emph{J. Chem. Phys.} 134:074109.

\bibitem{Jamadagni:ARCB:2011}
Jamadagni SN, Godawat R, Garde S
\newblock (2011) Hydrophobicity of proteins and interfaces: Insights from
  density fluctuations.
\newblock \emph{Annu Rev Chem Biomol Engg} 2:147--171.

\bibitem{Patel:PNAS:2011}
Patel AJ, {et~al.}
\newblock (2011) Extended surfaces modulate hydrophobic interactions of
  neighboring solutes.
\newblock \emph{Proc. Natl. Acad. Sci. U.S.A.} 108:17678 -- 17683.

\bibitem{Rotenberg:JACS:2011}
Rotenberg B, Patel AJ, Chandler D
\newblock (2011) Molecular explanation for why talc surfaces can be both
  hydrophilic and hydrophobic.
\newblock \emph{J. Am. Chem. Soc.} 133:20521 -- 20527.

\bibitem{Patel:JPCB:2012}
Patel AJ, {et~al.}
\newblock (2012) Sitting at the edge: How biomolecules use hydrophobicity to
  tune their interactions and function.
\newblock \emph{J. Phys. Chem. B} 116:2498 -- 2503.

\bibitem{remsing2013dissecting}
Remsing RC, Weeks JD
\newblock (2013) Dissecting hydrophobic hydration and association.
\newblock \emph{J Phys Chem B} 117:15479--15491.

\bibitem{Ashbaugh:JCP:2013}
Ashbaugh HS
\newblock (2013) Solvent cavitation under solvophobic confinement.
\newblock \emph{J Chem Phys} 139:064702.

\bibitem{Altabet:JCP:2014}
Altabet YE, Debenedetti PG
\newblock (2014) The role of material flexibility on the drying transition of
  water between hydrophobic objects: A thermodynamic analysis.
\newblock \emph{J. Chem. Phys.} 141:18C531.

\bibitem{Charlaix:PNAS:2012}
Guillemot L, Biben T, Galarneau A, Vigier G, Charlaix {\'E}
\newblock (2012) Activated drying in hydrophobic nanopores and the line tension
  of water.
\newblock \emph{Proc Natl Acad Sci USA} 109:19557--19562.

\bibitem{Miller:PNAS:2007}
Miller T, Vanden-Eijnden E, Chandler D
\newblock (2007) Solvent coarse-graining and the string method applied to the
  hydrophobic collapse of a hydrated chain.
\newblock \emph{Proc. Natl. Acad. Sci. U.S.A.} 104:14559--14564.

\bibitem{Willard:JPCB:2010}
Willard AP, Chandler D
\newblock (2010) Instantaneous liquid interfaces.
\newblock \emph{J. Phys. Chem. B} 114:1954--1958.

\bibitem{Vega:JCP:2007}
Vega C, De~Miguel E
\newblock (2007) Surface tension of the most popular models of water by using
  the test-area simulation method.
\newblock \emph{J Chem Phys} 126:154707.

\bibitem{Mittal:JCP:2012}
Mittal J, Hummer G
\newblock (2012) Pair diffusion, hydrodynamic interactions, and available
  volume in dense fluids.
\newblock \emph{J Chem Phys} 137:034110.

\bibitem{Li:JPCB:2012}
Li J, Morrone JA, Berne B
\newblock (2012) Are hydrodynamic interactions important in the kinetics of
  hydrophobic collapse?
\newblock \emph{J Phys Chem B} 116:11537--11544.

\bibitem{Mondal:PNAS:2013}
Mondal J, Morrone JA, Berne B
\newblock (2013) How hydrophobic drying forces impact the kinetics of molecular
  recognition.
\newblock \emph{Proc Natl Acad Sci USA} 110:13277--13282.

\bibitem{Setny:PNAS:2013}
Setny P, Baron R, Michael Kekenes-Huskey P, McCammon JA, Dzubiella J
\newblock (2013) Solvent fluctuations in hydrophobic cavity--ligand binding
  kinetics.
\newblock \emph{Proc Natl Acad Sci USA} 110:1197--1202.

\bibitem{Kumar:JCP:2011}
Kumar V, Sridhar S, Errington JR
\newblock (2011) Monte carlo simulation strategies for computing the wetting
  properties of fluids at geometrically rough surfaces.
\newblock \emph{J Chem Phys} 135:184702.

\bibitem{spce}
Berendsen HJC, Grigera JR, Straatsma TP
\newblock (1987) The missing term in effective pair potentials.
\newblock \emph{J. Phys. Chem.} 91:6269--6271.

\bibitem{Varilly:JPCB:2013}
Varilly P, Chandler D
\newblock (2013) Water evaporation: A transition path sampling study.
\newblock \emph{J Phys Chem B} 117:1419--1428.

\bibitem{Bussi:JCP:2007}
Bussi G, Donadio D, Parrinello M
\newblock (2007) Canonical sampling through velocity rescaling.
\newblock \emph{J. Chem. Phys.} 126:014101.

\bibitem{PME}
Essmann U, {et~al.}
\newblock (1995) A smooth particle mesh ewald method.
\newblock \emph{J. Chem. Phys.} 103:8577--8593.

\bibitem{SHAKE}
Ryckaert JP, Ciccotti G, Berendsen HJC
\newblock (1977) Numerical integration of the cartesian equations of motion of
  a system with constraints: molecular dynamics of n-alkanes.
\newblock \emph{J. Comp. Phys.} 23:327 -- 341.

\end{thebibliography}

\begin{thebibliography}{8}

\bibitem{Willard:JPCB:2010}
Willard,~A.~P.; Chandler,~D. {\em Instantaneous Liquid Interfaces}, J. Phys.
  Chem. B, {114}, 1954--1958 (2010).
  
\bibitem{marching_cubes}
Lorensen,~W.~E.; Cline,~H.~E. {\em Marching cubes: A high resolution 3D surface
  construction algorithm}, Computer Graphics, {21},
  163--169 (1987).
  
\bibitem{WHAM}
Ferrenberg,~A.~M.; Swendsen,~R.~H. {\em Optimized Monte Carlo data analysis},
  {Phys. Rev. Lett.}, {63}, 1195--1198 (1989).
  
\bibitem{UWHAM}
Tan,~Z.; Gallichio,~E.; Lapelosa,~M.; Levy,~R.~M. {\em Theory of binless multi-state
  free energy estimation with applications to protein-ligand binding}, {J.
  Chem. Phys.}, {136}, 144102 (2012).
  
\bibitem{MBAR}
Shirts,~M.~R.; Chodera,~J.~D. {\em Statistically optimal analysis of samples from
  multiple equilibrium states}, {J. Chem. Phys.}, {129}, 124105 (2008).

\bibitem{Sedlmeier:2012aa}
Sedlmeier,~F.; Netz,~R.~R. {\em The spontaneous curvature of the water-hydrophobe
  interface}, {J. Chem. Phys.}, {137}, 135102 (2012).

\bibitem{Suri:PRL:2014}
Vaikuntanathan,~S.; Geissler,~P.~L. {\em Putting water on a lattice: The importance
  of long wavelength density fluctuations in theories of hydrophobic and
  interfacial phenomena}, {Phys. Rev. Lett.}, {112}, 020603 (2014).

\bibitem{Sharma:2012aa}
Sharma,~S.; Debenedetti,~P.~G. {\em Evaporation rate of water in hydrophobic
  confinement} {Proc. Natl. Acad. Sci. U S A}, {109}, 4365--70 (2012).
\end{thebibliography}
